\documentclass[aps,prl,floatfix,twocolumn,superscriptaddress,showpacs,longbibliography,nofootinbib]{revtex4-2}

\usepackage{bibunits}
\usepackage{dcolumn}    
\usepackage{bm} 
\usepackage{graphicx}
\usepackage{amsmath}    
\usepackage{enumitem}
\usepackage{latexsym}
\usepackage{amsfonts}   
\usepackage{amssymb}
\usepackage{array}      
\usepackage{epsfig}
\usepackage{braket}
\usepackage{xcolor}
\usepackage[colorlinks=true,linkcolor=blue,urlcolor=blue,citecolor=blue,pdfusetitle]{hyperref}
\usepackage{hyperref}
\usepackage{pdfpages}
\usepackage{pgffor}
\makeatletter
\AtBeginDocument{\let\LS@rot\@undefined}
\makeatother

\newcommand{\ignore}[1]{}

\newcommand{\eq}{Eq.\,}
\newcommand{\eqs}{Eqs.\,}
\newcommand{\fig}{Fig.\,}

\newcommand{\cf} {cf.~}

\newcommand{\rref} {Ref.\,}
\newcommand{\rrefs} {Refs.\,}

\begin{document}

	\title{Vacancy-like dressed states in topological waveguide QED}

	\author{Luca Leonforte}
	\affiliation{Universit$\grave{a}$  degli Studi di Palermo, Dipartimento di Fisica e Chimica -- Emilio Segr$\grave{e}$, via Archirafi 36, I-90123 Palermo, Italy}

	\author{Angelo Carollo}
	\affiliation{Universit$\grave{a}$  degli Studi di Palermo, Dipartimento di Fisica e Chimica -- Emilio Segr$\grave{e}$, via Archirafi 36, I-90123 Palermo, Italy}
	\affiliation{Radiophysics Department, National Research Lobachevsky State University of Nizhni Novgorod, 23 Gagarin Avenue, Nizhni Novgorod 603950, Russia}

	\author{Francesco Ciccarello}
	\affiliation{Universit$\grave{a}$  degli Studi di Palermo, Dipartimento di Fisica e Chimica -- Emilio Segr$\grave{e}$, via Archirafi 36, I-90123 Palermo, Italy}
	\affiliation{NEST, Istituto Nanoscienze-CNR, Piazza S. Silvestro 12, 56127 Pisa, Italy}
	
	\date{\today}
	
	\begin{abstract}
		We identify a class of dressed atom-photon states forming
		at the same energy of the atom at any coupling strength. As a hallmark, their photonic component is an eigenstate of the bare photonic bath with a vacancy in place of the atom. The picture accommodates waveguide-QED phenomena where atoms behave as perfect mirrors, connecting in particular dressed bound states (BS) in the continuum or BIC with  geometrically-confined photonic modes.
		When applied to photonic lattices, the framework establishes a one-to-one correspondence between topologically-robust (TR) dressed states and TR photonic BS seeded by a vacancy.
		This is used to predict new classes of dressed BS in the photonic Creutz-ladder and Haldane models. In the latter case, states with non-zero local photon flux occur, where an atom is dressed by a photon orbiting around it. 
	\end{abstract}

	\maketitle
	
		Atom-photon dressed states are a basic concept of quantum electrodynamics (QED) \cite{CohenAP,harocheExploring2006}. A dressed {\it bound} state (BS), in particular, features a photonic cloud that remains localized close to the atom. Dramatic departure from spontaneous decay thus occurs, such as vacuum Rabi oscillations \cite{harocheExploring2006,raimondManipulating2001} or population trapping \cite{Bykov1975,John1994,Kofman1994,Lambropoulos2000a}.

	The interest in in-gap dressed BSs, in particular, has thrived in the last few years \cite{Longo2010,ShamoomPRA2013,Lombardo2014b,Douglas2015b,Calajo2016b,Tao2016,Gonzalez-Tudela2017a,Gonzalez-Tudela2017b,Liu2017a,Gonzalez-Tudela2018a,Tudela2018Quantum,Tudela2018anisotropic,Sundaresan2019,Sanchez-Burillo2019a,Roman-Roche2020}, prompting their experimental detection in various setups such as circuit QED \cite{Sundaresan2019} and cold atoms coupled to photonic crystal waveguides \cite{KimblePNAS2016} or optical lattices \cite{Krinner2018,Stewart2020}. A major appeal of dressed BSs is their ability to mediate dispersive many-body Hamiltonians \cite{ShamoomPRA2013,Douglas2015b,gonzalez2015subwavelength,Gonzalez-Tudela2018a,TaoNJP2018,Bello2019,garcia2019quantum,ZuecoSaw2020}.  Unlike cavity-QED schemes, these feature short-range, tunable interatomic couplings with promising applications in quantum technologies and many-body physics.
	
	In this Letter, we focus on a class of dressed states that can be both bound and unbound, which we dub ``Vacancy-like Dressed States" (VDS) for reasons that will become clear shortly. Their definition is simple: a VDS is a single-photon dressed state having exactly the same energy as the bare atom, irrespective of the coupling strength (under the rotating wave approximation). Familiar instances of dressed states, such as those arising in the Jaynes-Cummings model \cite{harocheExploring2006} and most of the in-gap BSs studied so far \cite{Douglas2015b}, are {\it not} VDS. While it might appear strange that the dressed-state energy can be insensitive to the coupling strength, in fact eigenstates with an analogous property recur in several fields such as quantum biology \cite{caruso2009} and dark states in atomic physics \cite{LP07}.
	
	As will be shown, the hallmark of VDS is that their photonic component is an eigenstate of the (bare) photonic bath with a vacancy on the atomic position (hence the name). Intuitively, the atom imposes a pointlike hard-wall boundary condition on the field and is then dressed by one of the resulting photonic eigenstates. This allows to embrace and reinterpret waveguide-QED phenomena \cite{RoyRMP17,LiaoPhyScr16,GuarXiv17} where atoms behave as perfect mirrors \cite{Shen2005,Zhou2008,Chang2007,Zhou2008,NoriPRAfabry,ciccarelloPRAgate,ChangNJP12,painter2019mirror}, in particular spotlighting the link between dressed BSs in the continuum (BIC) \cite{OrdonezPRA2006,TanakaPRB07,Longhi,TufarelliPRA13,GonzalezBallestroNJP13,TufarelliPRA14,RedchenkoPRA14,HoiNatPhy15,FacchiPRA16,Calajo2019,CornerPRR20,LonghiGiant2020} and photonic confined modes \cite{HsuNRM16}. When applied to topological photonic lattices, VDS prove essential for establishing general properties and occurrence criteria of topologically-protected dressed BSs, so far predicted and experimentally observed only in the photonic Su-Schrieffer-Heeger (SSH) model \cite{Bello2019,painter20}. Guided by this, new classes of topological dressed BSs are predicted in the photonic Creutz-ladder and Haldane models, highlighting potential applications and exotic properties such as persistent single-photon fluxes dressing the atom.

\begin{figure}
		\includegraphics[width=0.48 \textwidth]{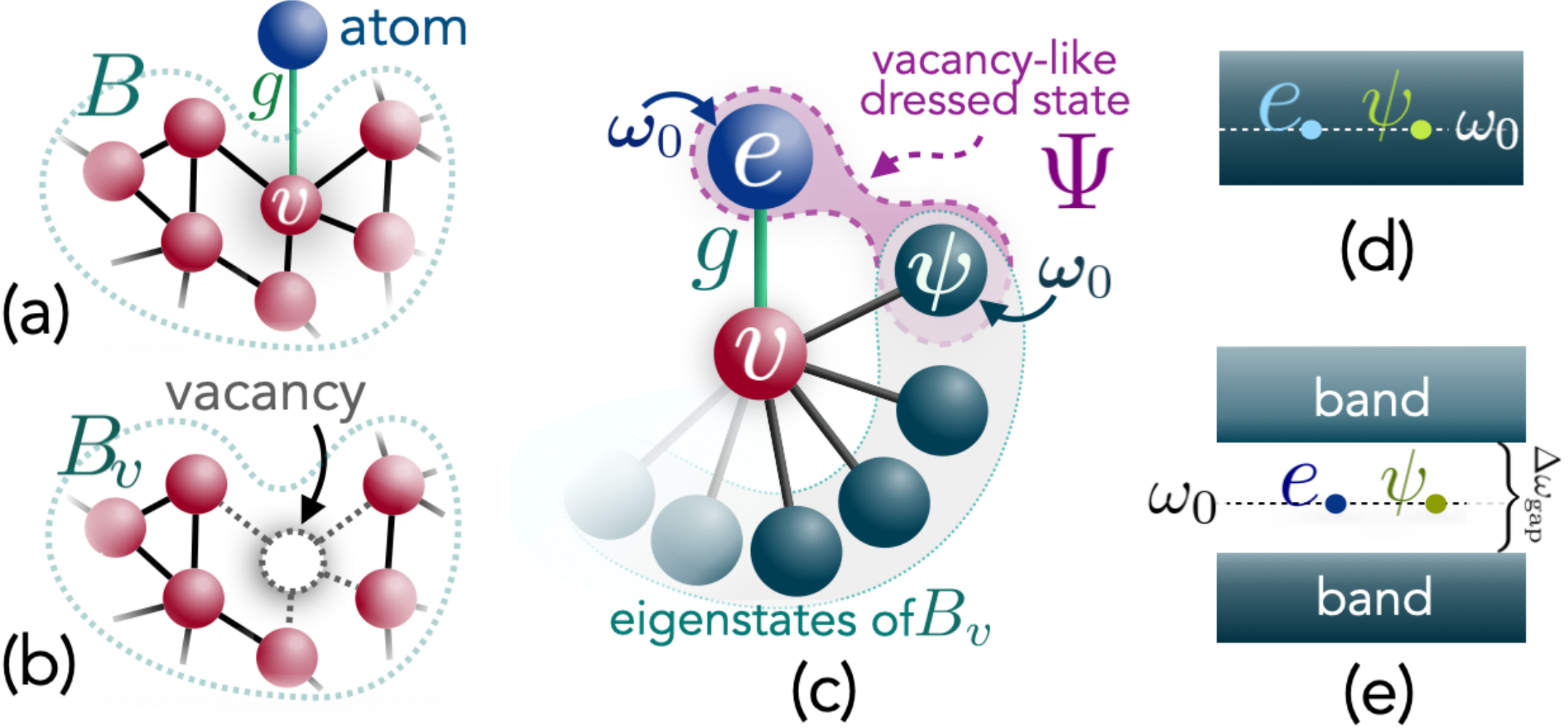}
		\caption{Basic framework of this work. (a): A generic photonic bath $B$ modeled as a network of coupled cavities. The atom is coupled to cavity $v$. (b): Same bath with cavity $v$ replaced by a vacancy, which we call $B_v$. (c): Representation of the $H$'s one-excitation sector: $\ket{v}$ is coupled to eigenstates of $H_{B_v}$ (among which $\ket{\psi}$) and the atom excited state $\ket{e}$. If $\ket{\psi}$ has the same energy as $\ket{e}$, a VDS $\ket{\Psi}$ exists. When $B$ has bands, both in-band VDS (d) and in-gap VDS (e) can occur.}\label{fig1}
	\end{figure}

	{\it Vacancy-like dressed states}.---Consider a general Hamiltonian model [see \fig\ref{fig1}(a)] describing a two-level (pseudo) atom with frequency $\omega_0$ weakly coupled to a structured photonic bath $B$ (field), the latter being an unspecified network of coupled bosonic modes (``cavities"). The Hamiltonian reads
	\begin{equation}
	H=\omega_0 \sigma_+\sigma_-+{H}_B+g\, ( b_{v}^\dag  \sigma_- + b_{v}  \sigma_+),\label{H}
	\end{equation}
	with 
	\begin{equation}
	H_B=\sum_i \omega_i  b_{i}^\dag  b_i+\sum_{i\neq j} \,J_{ij}  b_{i}^\dag  b_j\label{HB}
	\end{equation}
	being the bath free Hamiltonian ($J_{ji}=J_{ij}^*$). Here, $ b_i$ are bosonic ladder operators on $B$ fulfilling $[ b_{i}, b_j^\dag]=\delta_{ij}$, while $\sigma_-=\sigma_+^\dag=|g\rangle\langle e|$ are usual pseudospin ladder operators of the atom. 
	The atom is locally coupled to cavity $i=v$ (henceforth, at times referred to as the atom's ``position"). 
	
	A VDS $\ket{\Psi}$ is defined by
	\begin{equation}
	\ket{\Psi}\propto\varepsilon \ket{e}\ket{\rm vac} +\ket{g}\ket{\psi}\,,\label{Psi}
	\end{equation}
	fulfilling
	\begin{equation}
	H\ket{\Psi}=\omega_0 \ket{\Psi}\label{SE}\,
	\end{equation}
	with $\ket{\rm vac}$ the field vacuum and $\ket{\psi}=\sum_i \psi_i \ket{i}$ the photonic wavefunction (we denote single-photon states as $\ket{i}=b_i^\dag\! \ket{\rm vac}$).
	
	VDSs are single-photon dressed states with a {\it node} on the atom: $\psi_v=0$ (the converse holds as well). 
	This can be seen by projecting Eq. \eqref{SE} onto $\ket{e}$, yielding $\omega_{0}\varepsilon+g \psi_v =\omega_{0}\varepsilon$, hence $\psi_v=0$.
	Accordingly, if $B_v$ is the photonic bath obtained from $B$ by replacing cavity $v$ with a vacancy [see \fig\ref{fig1}(b)], the photon dressing the atom fully lives in $B_v$. It is then easily shown \cite{SM} that $\ket{\psi}$ is an eigenstate of $B_v$, again with energy $\omega_{0}$,
	\begin{equation}
	H_{B_v}\!\ket{\psi}=\omega_{0}\ket{\psi}\,,\label{SE2}
	\end{equation}
	where $ H_{B_v}$ (free Hamiltonian of $B_v$) is obtained from \eqref{HB} by restricting the sum to $i,j\neq v$. Also \cite{SM}, 
	\begin{equation}
	g\varepsilon+\langle v |H_B|\psi\rangle=0\,,\label{cond}
	\end{equation}
	where, explicitly, $\langle v |H_B|\psi\rangle=\sum_{i\neq v}J_{v,i }\psi_i$.
	Conversely, given $|\psi\rangle$ fulfilling Eq. \eqref{SE2}, the superposition of $\ket{e}$ and $\ket{\psi}$ defined by Eq. \eqref{cond} is a VDS.
	
	Thus a one-to-one mapping exists between VDS and single-photon eigenstates of the bare photonic bath with a vacancy in place of the atom (note that $\ket{\psi}$ is {\it not} an eigenstate of $ H_B$): Searching for VDSs in fact reduces to searching for {\it photonic normal modes in the presence of a vacancy}.
	
	We point out that, for each $\ket{\psi}$ fulfilling Eq. \eqref{SE2}, the existence of the VDS is guaranteed regardless of the coupling strength $g$ and bath structure. This is easily seen from the star-like structure of $H$ in the single-excitation sector, which is spanned by $\{\ket{e},\ket{i}\}$ (we conveniently introduce the compact notation $\ket{e}\ket{\rm vac}\rightarrow \ket{e}$, $\ket{i}\ket{\rm vac}\rightarrow \ket{i}$). Owing to the $\Lambda$ configuration with vertexes $\ket{v}$, $\ket{e}$ and $\ket{\psi}$ [see \fig\ref{fig1}(c)] it is clear that, when $\ket{e}$ and $\ket{\psi}$ have the same energy $\omega_{0}$, there {always} exists a superposition $\ket{\Psi}$ of them which, through destructive interference, is uncoupled from all other states (in formal analogy with, e.g., dark states \cite{LP07}). 
	
	In general, $\ket{\Psi}$ can be normalizable (i.e., a dressed BS in/out of the continuum) or not (i.e., unbound). Also, degeneracies can occur. When $\ket{\Psi}$ is {\it bound}, condition $\langle \Psi\ket{\Psi}=1$ and \eqref{cond} can be used to express it in the form
	\begin{equation}
	\ket{\Psi}=\cos\theta\,|e\rangle|{\rm vac}\rangle+e^{i \varphi} \sin\theta \, \ket{g}\ket{\psi}\,,\label{dressed}
	\end{equation}
	where
	\begin{align}
	\theta=\arctan{|\eta|}\,,\,\,\varphi=\arg{\eta}\,\,\,\,\,{\rm with}\,\,\eta=-\frac{g}{\langle v |H_B|\psi\rangle}\,\label{angle}
	\end{align}
	($\ket{\psi}$ fulfills $\langle \psi\ket{\psi}=1$ and Eq. \eqref{SE2}).
	
	\begin{figure*}
		\includegraphics[width=1.0 \textwidth]{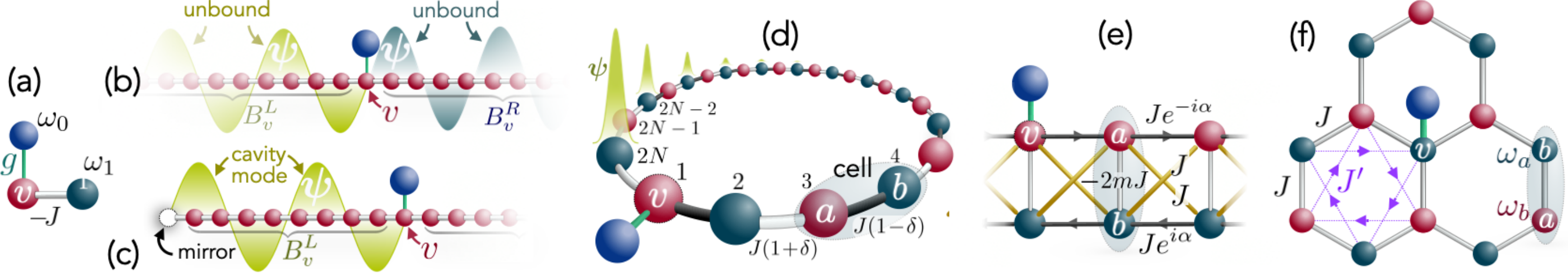}
		\caption{Instances of photonic baths where vacancy-like dressed states can occur: two cavities (a), an infinite waveguide (b), a semi-infinite waveguide (c), SSH model (d), Creutz-ladder model (e), Haldane model (f). In (d)-(e), $\omega_{a}=\omega_{b}=\omega_c$. In (f), $\omega_{a}=\omega_c\!+\! m J$, $\omega_{b}=\omega_c\!-\! m J$ and $J'=t e^{i\phi} J$.}\label{fig2}
	\end{figure*}
	VDS can be extended to continuous baths and many atoms~\cite{SM}.
	
	\textit{Two cavities}.---The simplest VDS occurs when $B$ is a pair of cavities ($v$ and $1$) coupled with strength $-J$ [see \fig\ref{fig2}(a)]. When $\omega_0=\omega_{1}$, \eqref{SE2} has the only solution $\ket{\psi}=\ket{1}$ yielding the VDS defined by $\tan \theta = g/J$,  $\varphi=0$. In all other dressed states the photon can be found at $v$. 
	Instead, for $\omega_0\neq \omega_{1}$, \eqref{SE2} has no solution and no VDS arises. Also, no VDS exists for $J=0$ (usual Jaynes-Cummings model \cite{harocheExploring2006}).

	\textit{Atom as a mirror}.---When $B$ is a 1D waveguide, where a vacancy is equivalent to a perfect {\it mirror}, VDSs formalize the known mirror-like behavior of atoms \cite{RoyRMP17}. 
	
	Let $B$ be an infinite waveguide (discretized for the sake of argument) [see \fig\ref{fig2}(b)] with $\omega_0$ well within the photonic band [see \fig\ref{fig1}(d)]. Then $B_v$ is the waveguide with a perfect mirror on the atom's location, i.e., $B_v=B^L_v \cup B^R_v$ with $B^L_v$ ($B^R_v$) the semi-infinite waveguide on the left (right) of $v$. Clearly, the eigenstates of $H_{B_v}$ are a continuum of sinusoidal, {\it unbound}, stationary waves with a node on $v$, each living either in $B^R_v$ or $B^L_v$ [see \fig\ref{fig2}(b)]. The pair at energy $\omega=\omega_{0}$ fulfill \eq\eqref{SE2}, thus two VDS exist. Each is a scattering state describing a left- or right-incoming photon of frequency $\omega_{0}$ fully reflected back from the atom, a major effect in waveguide QED
	\cite{Shen2005,ShenPRA09I,Chang2007} (see \rref\cite{SM} for details). 
	The one above is an instance of unbound VDS. 
	
	When the waveguide is semi-infinite [see \fig\ref{fig2}(c)], $B_v^L$ turns into a perfect cavity the related eigenstates being now discrete and {\it bound}, 
	each corresponding to a cavity protected mode with wavevector $k_m=m\pi/d$ ($m=1,2,...$) and frequency $\omega_{k_m}$. A {\it bound} VDS will thus arise when an $m$ exists such that $\omega_{k_m}=\omega_{0}$, i.e., $k_0 d=m \pi$ (with $k_0$ defined by $\omega_{k_0}=\omega_{0}$). Since the system is gapless, this VDS is a dressed BS {\it in the continuum} (BIC). Its explicit form is obtained from that of a textbook cavity mode by a direct application of \eqs\eqref{dressed}-\eqref{angle} \cite{SM}. A VDS for two atoms in an infinite waveguide is found likewise \cite{SM}.
	We thus retrieve a class of dressed BIC (or quantum BIC) \cite{OrdonezPRA2006,TanakaPRB07,Longhi,TufarelliPRA13,GonzalezBallestroNJP13,TufarelliPRA14,RedchenkoPRA14,HoiNatPhy15,FacchiPRA16,Calajo2019,CornerPRR20,LonghiGiant2020}: the VDS framework explicitly connects these quantum BIC to geometrically-confined photonic modes (corresponding respectively to $\ket{\Psi}$ and $\ket{\psi}$).
	
	
{\it  Topologically-robust (TR) dressed BS.}---By coupling an atom to a {translationally-invariant} photonic lattice possessing topological phases \cite{KaneRMP10,CarusottoPRL09,Altland1997}, an in-gap dressed BS can arise which is topologically robust (TR). To date, this was shown, both theoretically \cite{Bello2019} and experimentally \cite{painter20}, only for the SSH model [see \fig2(d)]. Criteria for occurrence of such states in a generic lattice and their general properties were currently unknown. The VDS picture fills this gap, as discussed next.

Let $B$ be a lattice (for now unspecified) under periodic boundary conditions BCs. The single-excitation spectrum of $H_B$ comprises continuous bands of unbound modes, separated by bandgaps. The total Hamiltonian $H$ [\cf\eq\eqref{H}] shares the same bands as $ H_B$ \cite{Kofman1994} each band now corresponding to a continuum of unbound {\it dressed} states. Additionally, in-gap dressed BSs -- at most one per bandgap -- generally occur \cite{economou2006green,Kofman1994,SM}. In particular, when $\omega_{0}$ lies within a bandgap 
a dressed BS can exist in the same bandgap (reducing to $\ket{e}$ for $g\rightarrow 0$). 

Similarly, a vacancy of a lattice $B$ can seed in-gap photonic BS of $H_{B}$, {at most one} within each finite (i.e., internal) bangap \cite{economou2006green,SM}. Then tuning the atom on resonance [see \fig\ref{fig1}(e)] with one such state, say $\ket{\psi}$, the corresponding VDS $\ket{\Psi}$ [\cf\eq\eqref{dressed}] will form. 
Now, based on the Altland-Zirnbauer classification of lattices~\cite{Schnyder2008,Kitaev2009a,Chiu2016}, if $B$ has the right symmetries then the vacancy can seed a TR midgap BS
of energy $\omega_\psi=\bar\omega_{c}$ with $\bar \omega_c$ the average frequency of the cavities (in general $\bar\omega_{c}={\rm Tr} H_B/M$ with $M$ the number of cavities)~\cite{Teo2010} . If so, coupling the atom to $B$ and tuning $\omega_{0}=\omega_\psi=\bar{\omega}_{c}$ will seed a VDS [\cf\eq\eqref{dressed}] inheriting the topological robustness of $\ket{\psi}$.
This rigorously formalizes the intuition, first appeared in \rref \cite{Bello2019}, that an atom can behave like an edge of a topological photonic bath, thus inducing topological BS.

The above provides a {\it general criterion} to find new classes of TR dressed BS (some instances are presented later on). 

Even more remarkably, using symmetry arguments one can show \cite{SM} that {\it any TR dressed BS {must be} a VDS}.




	{\it  Many-atom effective Hamiltonian.}--- A natural question is whether VDS are resilient to an imperfect setting of condition $\omega_0=\omega_\psi$ which their existence rely on. In \rref\cite{SM} this is shown to be the case for $g\ll \Delta\omega_{\rm gap}$ with $\Delta\omega_{\rm gap}$ the bandgap width.
	
	In this regime, when many atoms are present, it is known that in-gap dressed BS can mediate decoherence-free atom-atom interactions described by an effective Hamiltonian $H_{\rm eff}$ \cite{Douglas2015b,TaoNJP2018}. In the case of in-gap VDS, $H_{\rm eff} =\sum_{\nu\nu'}K_{\nu'\nu}\sigma_{\nu+}\sigma_{\nu'-}{+}{\rm H.c.}$, whose inter-atomic potential~\cite{SM}
\begin{equation}\label{Heff-MT}
	K_{\nu'\nu}=-\frac{g^2}{2\,\langle \nu' |H_B|\psi^{\nu}\rangle}\,\,\psi_{\nu'}^{\nu}\, 
\end{equation}
has just the same spatial profile as the photonic BS $\ket{\psi^{\nu}}$ arising when atom $\nu$ is replaced by a vacancy (in absence of all other atoms).
The interaction strength instead depends on how tightly connected is $\ket{\psi^\nu}$ to site $\ket{\nu'}$, this being measured by $\langle \nu' |H_B|\psi^{\nu}\rangle$ [\cf\eq\eqref{cond}]. Interestingly, apart from factor $g^2$, $K_{\nu'\nu}$ depends only on $H_B$ and $H_{B_v}$.
If the VDS is TR, so will be $H_{\rm eff}$.

Three instances of topological lattices follow.

{\it SSH model.}---The photonic SSH model \cite{SSH_PRL,SSH_PRB,Almeida2016,longhi2019landau} is the simplest 1D topological lattice [see \fig\ref{fig2}(d)]. The unit cell has two cavities, $a$ and $b$, both of frequency $\omega_{c}$, coupled with strength $J(1{-}\delta)$, where $|\delta |\le 1$, while the inter-cell coupling is $J(1{+}\delta)$. The total number of cavities is 2$N$ (even) with $N$ the number of cells. The $H_B$'s spectrum has two bands separated by a bandgap, centered at $\omega_{\rm mid}=\omega_c$, of width $\Delta \omega_{\rm gap}= 4 | \delta | J$. In this simple instance, $B_v$ is just an {\it open} SSH chain with an {\it odd} number of sites $2N{-}1$: this is well-known to exhibit (see, e.g., \rrefs\cite{Shin1997,Sirker2014}) a single in-gap TR edge state $\ket{\psi}$ of energy $\omega_{c}$ with non-zero amplitude only on sites of given parity.
If $v = a$, $\ket{\psi}$ is localized [see \fig\ref{fig2}(d)] close to the edge of $B_v$ on the right (left) of $v$ for $\delta>0$ ($\delta<0$) (right and left are swapped if $v=B$). 
Thus, for $\omega_{0}=\omega_c$, a corresponding TR VDS arises with a strongly asymmetric shape (``chiral BS" \cite{Bello2019,painter20}), which is worked out from the known form of $\ket{\psi}$  \cite{CiccarelloPRA2010} via a direct application of \eqref{dressed} \cite{SM}. Note that this dispenses with using the resolvent method \cite{CohenAP,Lambropoulos2000a}, by which this state was first found very recently \cite{Bello2019}.
Also note that, for $\delta=-1$, $B$ reduces to uncoupled pairs of cavities (dimers), linking this VDS to that for two cavities in \fig \ref{fig2}(a).
	\\
	\\
On a methodological ground, note that in 1D -- if $R \cdot d$ {\it edge states} \cite{KaneRMP10} of $B$ occur under {\it open} BCs (namely $B$ without a full cell) -- then a vacancy-induced BS always exists and can be inferred from these edge states. Here, integer $R$ is the interaction range of $B$ and $d$ the number of sites per cell (in the SSH model $R=1$ and $d=2$). This follows from a theorem proven in \rref\cite{SM}.	
	\\
	\\
	{\it Creutz-ladder (CL) model.}---Another 1D lattice with topological properties is the photonic CL model \cite{CreutzPRL99}, a circuit-QED implementation of which was recently put forward \cite{WilsonCreutz99}. The cell has again two cavities $a$ and $b$ each of frequency $\omega_{c}$ [see \fig\ref{fig2}(e)] with vertical (diagonal) coupling strength $-2m J$ ($J$), where $|m|\le 1$, and upper (lower) horizontal strength $J e^{-i \alpha}$ ($J e^{i \alpha}$).
	The bandgap is centered at $\omega_{\rm mid}=\omega_{c}-2m \cos\alpha J$, its width $\Delta\omega_{\rm gap}$ being the smallest of the four quantities $4 \delta_\pm J$ and $2 \left( \delta_+{+}\delta_-\pm 2 \cos \alpha \right) \!J$ with $\delta_\pm=| m {\pm} 1 | $. In particular, $\Delta\omega_{\rm gap}=0$ for $m=\pm1$.

	Using methods in \rrefs\cite{AlasePRB2017,CobaneraPRB218} combined with the aforementioned theorem for 1D lattices \cite{SM}, we find that, when $\Delta\omega_{\rm gap}>0$, $B_v$ admits a BS of energy $\omega_{\rm mid}$. This reads (we place the atom on site $a$ of cell $n=1$ and assume $N\gg 1$)
	\begin{equation}
	{\psi_{a_n}}=\tfrac{1}{2}\sqrt{1 {-} m^2}\left(e^{  i \alpha}  m^{n{-}2}  {+} e^{ -i \alpha} m^{N {-} n}\right)\,\,({\rm sites}\,a),\label{psi-cre}
	\end{equation}
	while $ {\psi_{b_n}}$ (sites $b$) is the same but $e^{ \pm i \alpha}{\rightarrow} -1$ (here, $n=2,...,N$; observe that cells on the left of $v$ are labeled by $N$, $N-1$,...). An analogous BS occurs for $v=b$ \cite{SM}. When $\omega_{0}=\omega_{\rm mid}$, a corresponding VDS is seeded being defined by [\cf\eqs\eqref{dressed}-\eqref{angle}] $\eta =\!- {g}/{(2J)}  e ^{i \frac{\alpha}{2}} \sin^{-2} \!\sqrt{1 {-} m^2}$. 
	Note that, unlike SSH, $|\psi_{a_n}|=|\psi_{b_n}|$. 
	Remarkably, for $\alpha=\pm \pi/2$ (such that $\omega_{\rm mid}=\bar \omega_c=\omega_{c}$) a topological phase occurs \cite{CreutzPRL99} ensuring that the above pair of edge states -- hence BS \eqref{psi-cre} and the associated VDS -- are topologically protected.
	
	In contrast to SSH, here no chirality manifests in the photon probability distribution since $|\psi_{j_n}|$ (for $j=a,b$) is mirror-symmetrical around $v$. The same holds for $\psi_{b_n}$ (phase included). Yet, $\psi_{a_n}\sim e^{ i \alpha}$ on the right of $v$ while $\psi_{a_n}\sim e^{- i \alpha}$ on the left. Thus, in the Creutz model, BS possess a chirality of {\it phase} (instead of modulus as in the SSH model). This is inherited by the corresponding VDS and thus by the following associated $H_{\rm eff}$. Plugging $\ket{\psi}$ into \eqref{Heff-MT} yields
	$K^{(aa)}_{n,n'}=\tfrac{g^2}{2m}\,e^{i \alpha}m^{n-n'}$ for two atoms sitting at cells $n$ and $n'$ both on sites $a$, while $K^{(bb)}_{n,n'}$ and $K^{(ab)}_{n,n'}$ are obtained from $K^{(aa)}_{n,n'}$ by replacing $\alpha$ with $-\alpha$ and $\pi$, respectively. This in particular allows to implement spin Hamiltonians with complex couplings \cite{ZuecoSaw2020} (e.g., placing all atoms on sites $a$), whose phase can be tuned via parameter $\alpha$ [see \fig\ref{fig2}(e)]. Moreover, for $\alpha=\pm\pi/2$ (see above) $H_{\rm eff}$ is topologically protected.

\begin{figure}
		\includegraphics[width=0.42 \textwidth]{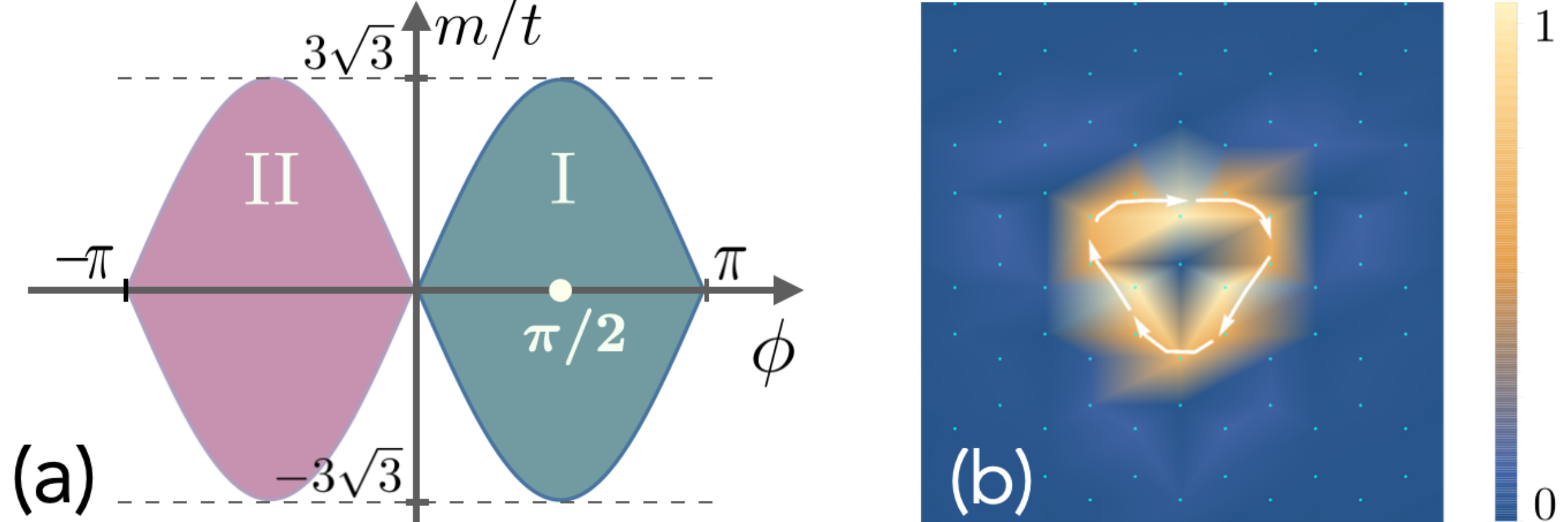}
		\caption{Haldane model. (a): Phase diagram. (b): Single-photon current density of the dressed bound state (VDS), i.e., of $\ket{\psi}$ (BS of $B_v$), for $m=0$, $\phi=\pi/2$ [white dot in panel (a)], $t=0.1$ and $v=a$ ($\Delta \omega_{\rm gap}\simeq 0.52 J$; for $g=0.01 J$, $\theta=0.057\,{\rm rad}$, $\varphi=0$). Plot (b) was obtained via exact numerical diagonalization  \cite{SM,Boykin2010} using a mesh of 30$\times$30 unit cells. The current density was rescaled to its maximum value $\simeq 0.26$. Photon probability density has similar shape and localization length.}\label{fig3}
	\end{figure}

	{\it Haldane model.}---The Haldane model is a prototypical 2D topological lattice~\cite{Haldane1988}, the first proposed to observe anomalous quantum Hall effect (QHE), whose photonic version \cite{polini2013} is considered next. Its honeycomb lattice [see \fig\ref{fig2}(f)] features a unit cell with two cavities ($a$ and $b$) of frequencies $\omega_c\pm m J$. Nearest-neighbour  (next-nearest-neighbour) cavity-cavity couplings are $J$ ($J'$) with $J'=J t e^{i\phi}$. The bandgap, centered at $\omega_{\rm mid}=\omega_{c}-3t\cos\phi J $, has width $\Delta\omega_{\rm gap}= ||m|-3\sqrt{3}t |\sin{\phi}|| J$. When $|m|<3\sqrt{3}t |\sin{\phi}|$ the model features two topological phases [named I and II in \fig\ref{fig3}(a)], witnessed under open BCs by a continuum of in-gap edge modes close to the lattice boundaries. These modes carry a stationary chiral current (as in the usual QHE \cite{QHE-PRL}).
	
	It can be shown \cite{SM} that a vacancy seeds an in-gap BS only within regions I-II.
	In particular, for $\phi=\pm \pi/2$ and $m=0$, the BS occurs at $\omega_{\rm mid}=\bar \omega_c=\omega_c$ (bandgap center) and is topologically robust~\cite{Teo2010,He2013}. 
Moreover, similarly to edge modes under open BCs, this BS features a chiral current density (CD) circulating around $v$ . A corresponding VDS thus arises for $\omega_{0}=\omega_{c}$ whose photonic component inherits analogous properties [see numerical instance in Fig.~\ref{fig3}(b)]. 
	We thus get that the the atom is dressed by a persistent single-photon current orbiting around it, a phenomenon with no 1D analogue. 
	
	We note that the $\phi$-$m/t$ plane [\cf\fig\ref{fig3}(a)] contains a whole set of points outside regions I-II having the same $\Delta \omega_{\rm gap}$ as \fig\ref{fig3}(b), where however (see above) no BS of $B_v$ occurs. Yet, in each point, for $\omega_{0}=\omega_{\rm mid}$ and $g$ small enough, an in-gap dressed BS (which is not a VDS) still arises. This also features a circulating CD, which is yet orders of magnitude weaker than the VDS in \fig\ref{fig3}(b) \cite{SM}. 
	
	{\it Conclusions.}---To sum up, we studied a class of dressed states, dubbed vacancy-like dressed states (VDSs), forming at the same energy as the atom. These are in one-to-one correspondence with normal modes of the bare photonic bath with a vacancy replacing the atom: if one among the latter modes has frequency matching the atom's then a VDS is seeded. Waveguide-QED phenomena where atoms behave as mirrors are naturally interpreted in terms of VDSs, based on which we explicitly linked dressed BIC to purely photonic bound modes.

	For photonic lattices, VDSs in fact provide a general criterion to find dressed BSs, and associated many-body Hamiltonians, inheriting topological properties (if any) of the bare photonic lattice. This was used to predict new classes of topological dressed BSs in the photonic Creutz-ladder and Haldane models. Either of these exhibits chiral properties. In the Creutz-ladder model, BSs show phase chirality (as opposed to modulus chirality in the SSH model). Haldane-model VDS instead feature a chiral single-photon current encircling the atom
	
VDSs play a central role in the emerging field of topological quantum optics where, as shown, any TR dressed state obtained by coupling an atom to a translationally invariant lattice must be a VDS. We note that this is different from coupling atoms to pre-existing TR modes of a lattice \cite{CiccarelloPRA2010,yao2013topologically,barik2018topological,longhi2019quantum,lemonde2019quantum}.
	
	Besides those studied here, we expect several other new classes of dressed BSs can be likewise unveiled (in particular the Haldane model shares some features worth being explored with the Harper-Hofstadter model, which
was very recently investigated in Ref. \cite{debernardis2020lightmatter}).
From a broader perspective, this work suggests a new beneficial link between quantum optics in structured baths and areas such as photonic BIC \cite{HsuNRM16}, topological photonics and condensed matter \cite{KaneRMP10,carusottoRMP,Mehrabad2019}.
	\\
	\\
	\begin{acknowledgments}
	We gratefully acknowledge fruitful discussions with A.González-Tudela and G. Calajò.
	We acknowledge support from MIUR through project PRIN Project 2017SRN-BRK QUSHIP. AC acknowledges support from the Government of the Russian Federation through Agreement No.~074-02-2018-330 (2).
		\end{acknowledgments}


	
	\bibliography{WQEDedge}
	\bibliographystyle{apsrev4-1}

	

\section*{Supplementary material}

# Supplemental Material for
# "Vacancy-like dressed states in topological waveguide QED"'

Luca Leonforte,[1] Angelo Carollo,[1,2] and Francesco Ciccarello[1,3]

[1] *Università degli Studi di Palermo, Dipartimento di Fisica e Chimica – Emilio Segrè, via Archirafi 36, I-90123 Palermo, Italy*
[2] *Radiophysics Department, National Research Lobachevsky State University of Nizhni Novgorod,
23 Gagarin Avenue, Nizhni Novgorod 603950, Russia*
[3] *NEST, Istituto Nanoscienze-CNR, Piazza S. Silvestro 12, 56127 Pisa, Italy*

(Dated: February 15, 2021)

## CONTENTS

This Supplemental Material presents technical proofs of some properties and theorems discussed in the main text. We note that most of Section SM5 deals with essentially known material, which is yet not easily found in explicit form in the literature. This is used to formulate a general necessary and sufficient condition for an in-gap dressed BS to exist.

**Notation for states.** Throughout this Supplemental Material, we use the same compact notation for states lying in the single-excitation subspace as that adopted in Fig. 1(c) of the main text. Thus, if the joint ground state is $|G\rangle = |g\rangle |\text{vac}\rangle$ with $|g\rangle$ the atom's ground state and $|\text{vac}\rangle$ the field vacuum, the single-excitation subspace is spanned by $|e\rangle = \sigma_+ |G\rangle$ (atom excited and no photons) and $\{|i\rangle\}$ with $|i\rangle = b_i^\dagger |G\rangle$ (one photon in the $i$th cavity with all the other cavities in the vacuum state and the atom in $|g\rangle$).

## SM1.   PROOF OF EQS. (5) AND (6)

We decompose Hamiltonian $H_B$ as

$$H_B = H_v + H_{B_v} + V_{v-B_v} , \qquad (S1)$$

with $H_v = \omega_v b_v^\dagger b_v$ the free Hamiltonian of cavity $v$ and

$$V_{v-B_v} = b_v^\dagger \sum_{i \neq v} J_{vi} b_i + \text{H.c.} \qquad (S2)$$

the coupling Hamiltonian between cavity $v$ and $B_v$ (bath with vacancy). Now note that

$$H |\psi\rangle = H_{B_v} |\psi\rangle + \sum_{i \neq v} J_{vi} \psi_i |v\rangle , \qquad (S3)$$

where we used decomposition (S1) and (see main text) $\psi_v = \langle v | \psi \rangle = 0$. On the other hand, using (3), Eq. (4) can be rearranged as

$$H |\psi\rangle = \omega_0 |\psi\rangle - \varepsilon (H - \omega_0) |e\rangle = \omega_0 |\psi\rangle - g\varepsilon |v\rangle . \qquad (S4)$$

Upon comparison with (S3), we get

$$H_{B_v} |\psi\rangle + \sum_{i \neq v} J_{vi} \psi_i |v\rangle = \omega_0 |\psi\rangle - g\varepsilon |v\rangle . \qquad (S5)$$

Projecting either side onto $|i \neq v\rangle$, we just end up with (5) in matrix form, hence (5) holds true. Projecting instead (S5) onto $|v\rangle$ we get Eq. (6).

## SM2.   VACANCY-LIKE DRESSED STATES FOR CONTINUOUS BATHS

Vacancy-like dressed states can be defined even if the bath is a continuous field, instead of a discrete set of modes as in the main text. Let $b(x)$ be the ladder operators of the bath B satisfying $[b(x), b(x')] = 0$ and $[b(x), b^\dagger(x')] = \delta(x-x')$. The interaction Hamiltonian becomes

$$H_I = g \int dx\, \delta(x - x_v) \left( b^\dagger(x) \sigma_- + \text{H.c.} \right) \qquad (S6)$$

with $x_v$ the atom location. A single-excitation eigenstate of the full Hamiltonian $H = \omega_0 \sigma_+ \sigma_- + H_B + H_I$ reads

$$|\Psi\rangle = \varepsilon |e\rangle + |\psi\rangle = \left( \varepsilon\, \sigma_+ + \int dx\, \psi(x) b^\dagger(x) \right) |\text{vac}\rangle \qquad (S7)$$

with $H |\Psi\rangle = \omega |\Psi\rangle$. The projections of $|\Psi\rangle$ on $|e\rangle$ and $|x\rangle = b^\dagger(x) |0\rangle$ yield, respectively,

$$\varepsilon\, \omega_0 + g \langle x_v | \psi \rangle = \varepsilon\, \omega , \qquad (S8)$$

$$\varepsilon g \langle x | x_v \rangle + \langle x | H_B | \psi \rangle = \omega \langle x | \psi \rangle . \qquad (S9)$$

When $\omega = \omega_0$, the eigenstate $|\Psi\rangle$ is a VDS. This definition, together with Eq. (S8) leads to the following necessary and sufficient condition for a VDS

$$\langle x_v|\psi\rangle = \psi(x_v) = 0\,, \tag{S10}$$

which is the continuous counterpart of the *node equation* $\psi_\nu = 0$ (see main text). By manipulating Eq. (S9) and using $\langle x|x_v\rangle = \delta(x - x_v)$, we derive the following constraint on the photonic component $|\psi\rangle$ of a generic eigenstate $|\Psi\rangle$

$$H_{B_v} |\psi\rangle = \left(H_B + \frac{\varepsilon g}{\psi(x_v)} |x_v\rangle\langle x_v|\right) |\psi\rangle = \omega |\psi\rangle\,. \tag{S11}$$

The above expression suggests that $|\psi\rangle$ is an eigenstate of the bath Hamiltonian $H_B$ with an extra point-like potential centered on $x = x_v$. For a VDS, this potential turns into a hard-wall condition, due to the node equation of Eq. (S10). This, together with the resonance condition $\omega = \omega_0$, turns Eq. (S11) into the continuous analogue of Eq. (5) in the main text.

### SM3. MANY-ATOM VACANCY-LIKE DRESSED STATES

The generalization of Hamiltonian (1) to $N_a$ atoms indexed by $\nu = 1, ..., N_a$ with the $\nu$th atom coupled to cavity $\nu$ reads

$$H = \omega_0 \sum_{\nu=1}^{N_a} \sigma_{\nu+}\sigma_{\nu-} + H_B + g \sum_{\nu=1}^{N_a} \left(b_\nu^\dagger \sigma_{\nu-} + \text{H.c.}\right)\,. \tag{S12}$$

The single-excitation subspace is spanned by $\{\{|e_\nu\rangle\}, \{|i\rangle\}\}$ with $|e_\nu\rangle = \sigma_{\nu+} |G\rangle$. A vacancy-like dressed state (VDS) is defined as

$$|\Psi\rangle = \sum_{\nu=1}^{N_a} \varepsilon_\nu |e_\nu\rangle + |\psi\rangle \tag{S13}$$

with $H |\Psi\rangle = \omega_0 |\Psi\rangle$. Projecting the latter equation onto $|\nu\rangle$ yields $\omega_0 \varepsilon_\nu + g\psi_\nu = \omega_0 \varepsilon_\nu$, hence $\psi_\nu = 0$ for any $\nu$.

Let $B_v$ be now the bare bath with cavities $\nu = 1, ..., N_a$ replaced by vacancies, whose free Hamiltonian $H_{B_v}$ is obtained from (2) by restricting the sum to $i, j \neq \{\nu\}$ (i.e., all indexes $i, j$ different from $\nu = 1, .., N_a$). To show that $H_{B_v}|\psi\rangle = \omega_0 |\psi\rangle$, we decompose $H_B$ as

$$H_B = H_v + H_{B_v} + V_{v-B_v}\,, \tag{S14}$$

with $H_v = \sum_{\nu=1}^{N_a} \omega_\nu b_\nu^\dagger b_\nu$ and

$$V_{v-B_v} = \sum_{\nu=1}^{N_a} \sum_{i \in B_v} J_{\nu,i} b_\nu^\dagger b_i + \text{H.c.}\,, \tag{S15}$$

where the second sum runs over all cavities of $B_v$.

Now note that

$$H |\psi\rangle = H_{B_v} |\psi\rangle + \sum_{\nu=1}^{N_a} \sum_{i \in B_v} J_{\nu,i} \psi_i |\nu\rangle\,, \tag{S16}$$

where we used decomposition (S14) and $\psi_\nu = 0$ for any $\nu$. On the other hand, using (S13), $H |\Psi\rangle = \omega_0 |\Psi\rangle$ can be rearranged as

$$H |\psi\rangle = \omega_0 |\psi\rangle - \sum_{\nu=1}^{N_a} \varepsilon_\nu (H - \omega_0) |e_\nu\rangle = \omega_0 |\psi\rangle - g \sum_{\nu=1}^{N_a} \varepsilon_\nu |\nu\rangle$$

Upon comparison with (S16), we get

$$H_{B_v}|\psi\rangle + \sum_{\nu=1}^{N_a} \sum_{i \in B_v} J_{\nu,i} \psi_i |\nu\rangle = \omega_0 |\psi\rangle - g \sum_{\nu=1}^{N_a} \varepsilon_\nu |\nu\rangle$$

Finally, projecting either side onto $|i\rangle$ with $i \in B_v$, we just end up with $H_{B_v}|\psi\rangle = \omega_0 |\psi\rangle$ in matrix form, completing the proof.

Projecting onto $\nu = 1, ..., N_a$ instead yields

$$g\varepsilon_\nu + \sum_{i \in B_v} J_{\nu,i}\psi_i = 0, \tag{S17}$$

which can be used to express the atomic amplitudes in terms of $|\psi\rangle$.

## SM4. ATOM AS A MIRROR

### SM4.1. Perfect reflection of a resonant photon

Consider the model in the main text [cf. Eq. (1)] in the case that $B$ is a discrete infinite waveguide modeled as a homogeneous coupled-cavity array with cavities located at positions $x_n = n\Delta x$, where integer $n$ for the present lattice coincides with the cell index. The free Hamiltonian of $B$ has the usual tight-binding form

$$H_B = \omega_c \sum_{n=-\infty}^{\infty} b_n^\dagger b_n - J \sum_{n=-\infty}^{\infty} (b_{n+1}^\dagger b_n + \text{H.c.}) \tag{S18}$$

with $\omega_c$ the frequency of each cavity and $-J$ the cavity-cavity coupling strength. The waveguide spectrum is $\omega_k = \omega_c - 2J\cos(k\Delta x)$, with $k \in ]-\pi/\Delta x, \pi/\Delta x]$ the wavevector whose corresponding group velocity is $v_k = 2J\Delta x \sin(k\Delta x)$.

Setting $v = 0$ (atom coupled to cavity $n = 0$), $B_v$ is the union of the semi-infinite arrays $B_v^L$ and $B_v^R$, respectively defined by $n \in ]-\infty, -1]$ and $n = 1, 2, ...$, as sketched in Fig. 2(b) of the main text. The normal modes of $B_v$ are thus

$$b_k = \sum_{n=-\infty}^{-1} \sin(kx_n) b_n, \quad b_k' = \sum_{n=1}^{\infty} \sin(kx_n) b_n \quad (0 \leq k \leq \pi/\Delta x) \tag{S19}$$

respectively corresponding to $B_v^L$ and $B_v^R$ (any normal mode of $B_v^{L(R)}$ is trivially also a normal mode of $B_v$ because $B_v^R$ and $B_v^L$ are uncoupled). The corresponding normal frequencies are $\omega_k = \omega_c - 2J\cos(k\Delta x)$ and $\omega_k' = \omega_c - 2J\cos(k\Delta x)$. Henceforth, we focus on $B_v^L$ (an analogous reasoning will apply to $B_v^R$).

The single-photon eigenstates of $H_{B_v}$ corresponding to modes $b_k$ are $|\psi_k\rangle$ with energy $\omega_k$ and wavefunction $\langle n |\psi_k\rangle = \sin(kx_n)$ for $n \leq -1$ and 0 otherwise. Hence, a VDS [cf. Eq. (3)] occurs for $|\psi\rangle = |\psi_{k_0}\rangle$, with a wave vector $k_0$ fullfilling $\omega_{k=k_0} = \omega_0$. Noting that we can write $\langle n|\psi\rangle \propto e^{ik_0 x_n} + re^{-ik_0 x_n}$ with $r = -1$, we see that the VDS is a scattering state describing a left-incoming photon fully reflected back from the atom as if this were a perfect mirror. Condition (6) in this case simply reads $g\varepsilon - J\sin(k_0\Delta x) = 0$, hence $\varepsilon = J\sin(k_0\Delta x)/g \propto v_{k_0}/g$, matching known results obtained via scattering theory (see e.g. Ref. [1]).

### SM4.2. Dressed bound states (BS) in the continuum (BIC): one atom in a semi-infinite waveguide

The previous infinite waveguide is now replaced by a semi-infinite waveguide made out of cavities $n = 1, 2, ...$, hence in each sum of Hamiltonian (S18) $n$ now starts from 1 (equivalently, one can think of the right half of an infinite waveguide with a perfect mirror on site $n = 0$). Placing the atom at site $d$ (thus $v = d$), $B_v$ [see Fig. 2(c) of the main text] is the union of the *finite* array $n = 1, 2, ..., d-1$ ($B_v^L$) and the semi-infinite lattice $n = d+1, d+2, ...$ (dubbed $B_v^R$). The normal frequencies of $B_v^R$ are the same as in the infinite-waveguide case with normal modes $b_k'$ [cf. Eq. (S19)] now displaced by the amount $v$. These are at the same time a (continuous) subset of normal frequencies and normal modes of $B_v$ (since $B_v^L$ and $B_v^R$ are disjoint). The remaining frequencies and normal modes of $B_v$ are those of $B_v^L$ (discrete). These are $\omega_{k_m} = \omega_c - 2J\cos k_m$, with $k_m = m\pi/d$ and $m = 1, 2, ..., d$, and

$$b_{k_m} = \sqrt{\tfrac{2}{d}} \sum_{n=1}^{d-1} \sin(k_m n) b_n. \tag{S20}$$

The corresponding single-photon *bound* eigenstates are $|\psi_{k_m}\rangle$, with $\langle n|\psi_{k_m}\rangle = \sqrt{2/d}\,\sin(k_m n)$ for $1 \leq n \leq d-1$ and $\langle n|\psi_{k_m}\rangle = 0$ for $n \geq d$, and energies $\omega_{k_m}$. A VDS (3) arises when one of these energies resonates with the atom,

i.e., there exists a value of $m$ such that $\omega_{k_m} = \omega_0$. By defining $k_0$ such that $\omega_{k=k_0} = \omega_0$, the VDS condition can be expressed in terms of wavevectors simply as $k_m = k_0$, that is

$$k_0 d = m\pi. \tag{S21}$$

Thus, if $m$ fulfills (S21), we set $|\psi\rangle = |\psi_{k_m}\rangle$. Since $|\psi\rangle$ is normalized, the corresponding VDS [cf. Eq. (7)] is a BS in the continuum or BIC (the atom frequency lies within the photonic band). To get the mixing angle $\theta$ [cf. Eq. (8) where $\varphi = 0$ in this case], we use that

$$\langle d-1 | \psi \rangle = \psi_{d-1} = \sqrt{\tfrac{2}{d}} \sin[k_m(d-1)] = \sqrt{\tfrac{2}{d}} (-1)^{m+1} \sin k_m = \sqrt{\tfrac{2}{d}} (-1)^{m+1} \tfrac{v_0}{2J}, \tag{S22}$$

where we set $v_0 = v_{k_m}$ (recall that $v_k = 2J \sin k$). Hence, $\theta = (-1)^{m+1} \arctan[\sqrt{2d}\,(g/v_0)]$, which once plugged into (7) yields the dressed BIC [2–4]

$$|\Psi\rangle = \frac{1}{\sqrt{1+\frac{\Gamma\tau}{2}}} \left( |e\rangle + (-1)^{m+1} \sqrt{\tfrac{2\Gamma}{v_0}} \sum_{n=1}^{d-1} \sin(k_m n) |n\rangle \right), \tag{S23}$$

where we introduced the decay rate $\Gamma = 2g^2/v_0$ and time delay $\tau = 2d/v_0$.

### SM4.3. Dressed BIC: two atoms in an infinite waveguide

A two-atom dressed BIC closely related to the previous one occurs in a waveguide, this time infinite. Let atoms 1 and 2 be coupled to cavities $n = 0$ and $n = d$, respectively. The setup is obtained from that in Fig. 2(c) of the main text by adding cavities $n = -\infty, ..., 0$ to the waveguide with cavity $n = 0$ ($n = d$) coupled to atom 1 (2). Domain $B_v$ (see Section SM3) is now the union of two semi-infinite waveguides (comprising sites $n < 0$ and $n > d$, respectively) plus the same cavity as in the previous section (i.e, the finite set of sites $n = 1, 2, ..., d$). As in the previous case, modes (S20) and $\omega_{k_m}$ are thus bound normal modes and normal frequencies of $B_v$. A two-atom bound VDS (S13) will thus arise with $|\psi\rangle \propto |\psi_{k_m}\rangle$ provided that $m$ fulfills condition (S21) [note that in (S13) $|\psi\rangle$ is not normalized]. Analogously to (S22),

$$\langle 1 | \psi_{k_m} \rangle = \sqrt{\tfrac{2}{d}} \tfrac{v_0}{2J}, \quad \langle d-1 | \psi_{k_m} \rangle = \sqrt{\tfrac{2}{d}} (-1)^{m+1} \tfrac{v_0}{2J} \tag{S24}$$

Thus, using Eq. (S17),

$$\varepsilon_1 = \tfrac{J}{g} \langle 1 | \psi_{k_m} \rangle = \sqrt{\tfrac{2}{d}} \tfrac{v_0}{2g}, \quad \varepsilon_2 = \tfrac{J}{g} \langle d-1 | \psi_{k_m} \rangle = \sqrt{\tfrac{2}{d}} (-1)^{m+1} \tfrac{v_0}{2g}, \tag{S25}$$

entailing $\varepsilon_1 = (-1)^{m+1} \varepsilon_2$. Therefore, through Eq. (S13), we get that the (unnormalized) VDS corresponding to $|\psi_{k_m}\rangle$ is

$$|\Psi\rangle = \tfrac{v_0}{g\sqrt{d}} |\Phi^{\pm}\rangle + |\psi_{k_m}\rangle, \tag{S26}$$

where $|\Phi^{\pm}\rangle = \tfrac{1}{\sqrt{2}}(|e_1\rangle \pm |e_2\rangle)$ and the plus (minus) sign holds for odd (even) $m$. Upon normalization, we end up with

$$|\Psi\rangle = \frac{1}{\sqrt{1+\frac{\Gamma\tau}{4}}} \left( |\Phi^{\pm}\rangle + \sqrt{\tfrac{\Gamma}{v_0}} \sum_{n=1}^{d-1} \sin(k_m n) |n\rangle \right),$$

matching the known expression of the two-atom dressed BIC obtained with other methods (see e.g. Refs. [4, 5]).

### SM4.4. Continuous waveguide

The above conclusions hold for a discrete waveguide (cavity array) but can be extended to a continuous waveguide with linear dispersion law using the results of Section SM2. The free Hamiltonian of $B$ is turned into

$$H_B = (\omega_c - v_g \bar{k}) \sum_{\alpha=R,L} \int_{-\infty}^{\infty} dx\, b_\alpha^\dagger(x) b_\alpha(x) - iv_g \int_{-\infty}^{\infty} dx\, b_R^\dagger(x) \partial_x b_R(x) + iv_g \int_{-\infty}^{\infty} dx\, b_L^\dagger(x) \partial_x b_L(x), \tag{S27}$$

with $v_g$ is the group velocity, while $\pm \bar{k}$ such that $\omega_c = \omega_{\pm \bar{k}}$ are the wavevectors around which we linearize the dispersion law $\omega_k$. Here, $b_R(x)$ are a set of bosonic continuous ladder operators fulfilling $[b_R(x), b_R^\dagger(x')] = \delta(x-x')$ [and similarly for left-going operators $b_L(x)$], where right-going operators (having subscript $R$) commute with left-going ones (subscript $L$).

The interaction Hamiltonian is given by (S6) of Section SM2 under the formal replacement $b(x) \to b_R(x) + b_L(x)$. Eqs. (S8) and (S9) can be applied by formally replacing $|x\rangle$ with $(b_R^\dagger(x) + b_L^\dagger(x))|\text{vac}\rangle$ and analogously for $|x_v\rangle$ (likewise for the corresponding bras).

### SM4.5. Perfect reflection of a resonant photon in a continuous waveguide

Setting $x_v = 0$ (atom at the origin), an (unnormalized) eigenstate of $B_v^R$, namely the semi-infinite waveguide $x > 0$, reads

$$|\psi\rangle = \int_0^\infty dx \, \frac{1}{2i} \left( e^{ik_0 x} b_R^\dagger(x) - e^{-ik_0 x} b_L^\dagger(x) \right) |\text{vac}\rangle , \qquad (S28)$$

which describes a right-incoming photon of wavevector $k_0$ fully reflected back from the atom. Hence,

$$\langle x | H_B | \psi \rangle = \langle \text{vac} | b_R(x) H_B | \psi \rangle + \langle \text{vac} | b_L(x) H_B | \psi \rangle =$$
$$= \int_{-\infty}^{+\infty} dy \int_0^{+\infty} dz \left[ (\omega_c - v_g \bar{k}) \sin(k_0 z) \delta(x-y) \delta(y-z) - v_g \cos(k_0 z) \delta(x-y) \partial_y \delta(y-z) \right]$$
$$= \theta(x)(\omega_c - v_g \bar{k}) \sin(k_0 x) + v_g k_0 \theta(x) \sin(k_0 x) - v_g \delta(x) \cos(k_0 x) .$$

By plugging this into Eq. (S9) one can show that a photon wavevector $k_0$ fulfilling $\omega_0 = \omega_c - v_g(\bar{k} - k_0)$ leads to an unbound VDS with $\varepsilon \propto \frac{v_g}{g}$, in line with the discrete waveguide of Section SM4.1.

### SM4.6. Dressed BICs in a continuous waveguide

For a continuous semi-infinite waveguide, $H_B$ is analogous to (S27) except that the field is now constrained to live on $x \geq 0$ with a node at $x = 0$. In the case of one atom placed at $x_v = d$, the single-photon eigenstates of $B_v^L$, namely the continuous counterpart of (S20), read

$$|\psi_{k_m}\rangle = \sqrt{\frac{2}{d}} \int_0^d dx \left( \frac{e^{ik_m x} b_R^\dagger(x) - e^{-ik_m x} b_L^\dagger(x)}{2i} \right) |\text{vac}\rangle \qquad (S29)$$

each having energy $\omega_{k_m} = \omega_c - v_g(\bar{k} - k_m)$, where $k_m = m\pi/d$ and $m = 1, 2, ...$

After some calculations, the matrix element entering Eq. (S9) is worked out as

$$\langle x | H_B | \psi_{k_m} \rangle = \langle \text{vac} | b_R(x) H_B | \psi_{k_m} \rangle + \langle \text{vac} | b_L(x) H_B | \psi_{k_m} \rangle$$
$$= \sqrt{\tfrac{2}{d}} \left[ (\omega_c + v_g(k - \bar{k})) \sin(kx) \theta(x) \theta(d-x) + v_g \delta(x-d) \cos(kd) - v_g \delta(x) \theta(0) \right] . \qquad (S30)$$

Thus, based on Eq. (S9), a VDS exists whenever there is a $k_m$ such that

$$\omega_0 - \omega_c = v_g(k_m - \bar{k}), \qquad (S31)$$

the corresponding mixing angle being given by $\theta = (-1)^{m+1} \arctan[\sqrt{d/2}\,(2g/v_g)]$. Accordingly, the full VDS can be written as [4]

$$|\Psi\rangle = \frac{1}{\sqrt{1 + \frac{\Gamma \tau}{2}}} \left( |e\rangle + (-1)^{m+1} \sqrt{\frac{\Gamma \tau}{2}} |\psi_{k_m}\rangle \right) \qquad (S32)$$

with $\Gamma = \frac{2g^2}{v_g}$ and $\tau = \frac{2d}{v_g}$.

The continuous version of the two-atom dressed BIC in Section SM4.3 is derived likewise.

## SM5. VACANCY IN A TRANSLATIONALLY-INVARIANT LATTICE: UNIQUENESS OF BS AND GENERAL CONDITION FOR HAVING A VDS

Here, we will show that *in a translationally-invariant lattice with a single vacancy* (i.e., $B_v$ with $B$ a translationally-invariant lattice) *there is at most one non-degenerate bound state (BS) in each internal bandgap*.

Consider a generic translationally-invariant $D$-dimensional lattice ($D = 1, 2, 3$) with finite-range interactions and $d$ sites in each unit cell. A vacancy on site $\alpha_v$ in the cell $n_v$ breaks translational invariance transforming the lattice Hamiltonian as $H_B \to H_{B_v} = H_B + H_1$ with the perturbation Hamiltonian defined by

$$H_1 = \epsilon \ket{n_v, \alpha_v} \bra{n_v, \alpha_v} \quad \text{for } \epsilon \to \infty. \tag{S33}$$

Consider the Green functions of the unperturbed and perturbed Hamiltonians, respectively defined as $G_B(z) = (z - H_B)^{-1}$ and $G_{B_v}(z) = (z - H_{B_v})^{-1}$, which fulfill [6]

$$G_{B_v}(z) = G_B(z) + G_B(z) T(z) G_B(z) \tag{S34}$$

with

$$T(z) = H_1 \frac{1}{\mathbb{1} - G_B(z) H_1} = \frac{\ket{n_v, \alpha_v}\bra{n_v, \alpha_v}}{1/\epsilon - \langle G_B(z) \rangle_v} \tag{S35}$$

with $\langle \ldots \rangle_v = \bra{n_v, \alpha_v} \ldots \ket{n_v, \alpha_v}$.

The bound states of $H_{B_v}$ correspond to the poles $z = \omega_p$ of $G_{B_v}(z)$, which are the solutions of the equation $\langle G_B(\omega_p) \rangle_v = 1/\epsilon$ [6], which for $\epsilon \to \infty$ (vacancy) simply reduces to

$$\langle G_B(\omega_p) \rangle_v = 0. \tag{S36}$$

Themlatter cannot be satisfied by an $\omega_p$ inside a band of $H_B$ since within each band $\langle G_B(\omega) \rangle_v$ has a non-zero imaginary part (this being proportional to the density of states [6]).

For $z = \omega \in \mathbb{R}$ outside of bands of $H_B$, $G_B$ is a Hermitian operator, analytical function of $z$, and $\langle G'_B(\omega) \rangle_v = -\langle (\omega - H_B)^{-2} \rangle_v = -\langle G_B(\omega)^2 \rangle_v \leq 0$ ($G'$ is the derivative with respect to $z$). This shows that, in each given bandgap (including the outermost semi-infinite bandgaps), $\langle G_B(\omega) \rangle_v$ *is a monotonic function of $\omega$*. Therefore, there is at most one solution of Eq. (S36) within each energy interval where $\langle G_B(\omega) \rangle_v$ is a continuous function of $\omega$, i.e., within each finite and semi-infinite bandgap.

However, notice that $\lim_{\omega \to \pm \infty} G_B(\omega) = 0$, hence there is no finite value of $\omega$ satisfying Eq. (S36) in the two external semi-infinite bandgaps (above and below the uppermost and lowermost bands). Therefore Eq. (S36) can only be fulfilled within an *internal* finite bandgap.

Finally, we derive the degeneracy $\delta_p$ of the BS of $H_{B_v}$ for a given solution $\omega_p$. The projector onto the eigenspace of $H_{B_v}$ corresponding to eigenvalue $\omega_p$ is the residue of $G_{B_v}(z)$ at the pole $z = \omega_p$. By plugging the expansion $\langle G_B(z) \rangle_v = 1/\epsilon + \langle G'_B(\omega_p) \rangle_v (z - \omega_p) + o(z - \omega_p)$ into Eq. (S34) and using Eq. (S35) one gets

$$\delta_p = \text{Tr}\{\text{Res}_{\omega_p}[G_{B_v}]\} = -\text{Tr}\left\{ \frac{G_B(\omega_p) \ket{n_v, \alpha_v}\bra{n_v, \alpha_v} G_B(\omega_p)}{\langle G'_B(\omega_p) \rangle_v} \right\} = -\frac{\langle G_B^2(\omega_p) \rangle_v}{\langle G'_B(\omega_p) \rangle_v} = 1. \tag{S37}$$

Thus each solution of Eq. (S36) corresponds to a non-degenerate BS.

Combining the above with the main text discussion about VDS, it turns out that a general necessary and sufficient condition for an in-gap VDS to arise is

$$\langle G_B(\omega_0) \rangle_v = 0 \tag{S38}$$

with $\omega_0$ the atomic transition frequency.

## SM6. ALL TOPOLOGICALLY-ROBUST DRESSED STATES ARE VDS

In this section, we will show that a *dressed BS seeded by an atom coupled to a symmetry-protected topological photonic lattice is topologically robust if and only if it is a VDS*.

Here, *topologically robust* means that both the existence and energy of the state are unaffected by any perturbation preserving the system's internal symmetries. To better clarify this concept, we first briefly review some known theory on symmetry-protected topological lattices [7, 8].

### SM6.1. Zero-dimensional defects and topologically-robust BS

According to the so called "ten-fold way" or Altland-Zirnbauer (AZ) classification of lattices [9–11], any translationally invariant lattice of dimension $d$ belongs to one of ten classes. These classes are identified by the occurrence or the absence of three different internal (i.e., non-spatial) symmetries: time-reversal ($\mathcal{T}$), particle-hole ($\mathcal{C}$) and chiral symmetry ($\mathcal{S}$). By introducing a defect of dimension $d_{\text{defect}}$ into a lattice (a point-like defect has $d_{\text{defect}} = 0$, a line defect $d_{\text{defect}} = 1$ etc.), a TR BS localized around the defect can appear only if the lattice admits a suitable non-trivial topological invariant. Weather or not this is the case depends only on the AZ class of the lattice and $d_{\text{defect}}$ [8].

Here, we are interested in TR BS arising when an atom is locally coupled to a translationally-invariant photonic lattice [cf. Eq. (1) in the main text]. Note that the atom can be regarded as a point-like defect of the lattice such that $d_{\text{defect}} = 0$. In this case, as shown in Fig. S1(a), the "ten-fold way" predicts that only five out of the ten AZ-classes can possess a non-trivial topological invariant: AIII, BDI, D, DIII and CII classes (using the standard nomenclature) [8]. As a key point, note that each of these five classes features at least one of the two symmetries $\mathcal{C}$ and $\mathcal{S}$ [9, 11]. For a system (not necessarily translationally invariant) with (single-particle) Hamiltonian $H$, these are defined by

$$\mathcal{C}: \quad U_C^\dagger H^T U_C = -H, \tag{S39}$$

$$\mathcal{S}: \quad U_S^\dagger H U_S = -H \tag{S40}$$

for some unitary matrices $U_C$ and $U_S$.

We show next that when either $\mathcal{C}$ or $\mathcal{S}$ are present then a BS occurring in a central bandgap must be TR. First, recall that the spectrum of an operator is invariant under transposition and unitary transformations. Thus, any $H$ fulfilling (S39) or (S40) must have an energy spectrum symmetric around zero [see Fig. S1(b)]: each eigenstate with positive energy must be paired with another eigenstate having the opposite energy. As a consequence, any (internal) bandgap containing zero must be just centered at zero energy [see central bandgap in Fig. S1(b)]. Most importantly, if a *single* (i.e, non-degenerate) BS occurs within such central bandgap this state cannot be paired with any other, hence its energy must be exactly zero. Remarkably, this conclusion holds unchanged in the presence of any perturbation not affecting Eq. (S39) or Eq. (S40)). Furthermore, for reasons related to the presence of a topological invariant, the existence of this BS is robust as long as the bandgap remains open [7, 12]. To sum up, both the existence and energy of such BS are insensitive to perturbations preserving the system's symmetries, namely the BS is TR according to the definition given at the beginning of this section.

### SM6.2. Atom coupled to a photonic lattice

We next apply the above argument to the specific case of an atom coupled to a translationally-invariant photonic lattice. The single-excitation total Hamiltonian [Eq. (1) in the main text] can be written as

$$H = (\omega_0 - \bar{\omega}_c) |e\rangle \langle e| + H_B + (g |e\rangle \langle v| + \text{H.c.}). \tag{S41}$$

Here, $\bar{\omega}_c$ is a zero-energy reference which is set in such a way that $\text{Tr}(H_B) = 0$. If all the cavities have the same bare frequency $\omega_c$, namely $\omega_i = \omega_c$ for any $i$, then $\bar{\omega}_c = \omega_c$. Assume now that $H_B$ (free Hamiltonian of the photonic lattice) has the right symmetries to seed a TR BS around a point-like defect. This means (see above) that its bare Hamiltonian $H_B$ fulfils (S39) for some unitary $U_C$ or (S40) for some unitary $U_S$.

We prove next that $\omega_0 = \bar{\omega}_c$ is a necessary and sufficient condition in order for the total Hamiltonian $H$ to fulfill the same internal symmetry of $H_B$. Namely, if $H_B$ enjoys the particle-hole symmetry $\mathcal{C}$ (chiral symmetry $\mathcal{S}$) then $H$ fulfils (S39) [(S40)] for some suitable unitary $\tilde{U}_{C(S)}$ iff $\omega_0 = \bar{\omega}_c$.

Let us then construct the operator $\tilde{U}_s$ (with $s = C, S$) from unitary $U_s$. Due to the uniqueness of $U_s$ [7], the matrix elements of $\tilde{U}_s$ must coincide with those of $U_s$ on the photonic degrees of freedom, namely

$$\tilde{U}_s = \alpha |e\rangle \langle e| + \sum_i (\beta_i |e\rangle \langle i| + \gamma_i |i\rangle \langle e|) + U_s \tag{S42}$$

where $\alpha$, $\beta_i$ and $\gamma_i$ are unknown costants ($i$ labels the lattice sites). The unitarity condition for $U_s$ and $\tilde{U}_s$ implies that $|\alpha| = 1$ and $\beta_i = \gamma_i = 0$ $\forall i$, thus yielding

$$\tilde{U}_s = e^{i\phi_s} |e\rangle \langle e| + U_s, \tag{S43}$$

(a)

FIG. S1. (a): Periodic table of topological lattices in the presence of a zero-dimensional defect ($d_{\text{defect}} = 0$). The left-most column (A, AIII, ...) reports the ten symmetry classes of the lattice single-particle Hamiltonian. Each is defined based on the presence/absence of three symmetries: time-reversal ($\mathcal{T}$), particle-hole ($\mathcal{C}$), and chiral ($\mathcal{S}$) symmetry (0 if the symmetry is absent, $\pm 1$ if present; see Ref. [7] for more details). The entries on the right-most column, in particular, denote the presence/absence of a topological invariant associated with a 0-dimensional defect: "0" if absent and '$\mathbb{Z}$", "$\mathbb{Z}_2$" or "$2\mathbb{Z}$" if present [7]. (b) Example of an energy spectrum of a lattice fulfilling (S39) or (S40). Each state with energy $E$ is paired with another one having energy $-E$. Hence, a bandgap across $E = 0$ must be just centered at $E = 0$. Importantly, if within this central bandgap a single BS occurs then its energy (see red line) must be exactly $E = 0$. The last statement holds even in the presence of a perturbation such that (S39) [(S40)] is still fulfilled for some unitary $U_C$ ($U_S$).

where $-\pi < \phi_s \leq \pi$ is a phase to be determined. Plugging now (S41) and (S43) into the particle-hole symmetry relation (S39) (with $\tilde{U}_C$ in place of $U_C$) leads to the following conditions

$$\phi_C = \chi_C + \pi, \qquad \omega_0 = \bar{\omega}_c. \tag{S44}$$

Here, we set $\langle v|U_C|v\rangle = e^{i\chi_C}$ (which defines phase $\chi_C$) taking advantage from the fact that $U_s$ is just a *diagonal* operator in the $\{|i\rangle\}$ basis. The last property holds because $\mathcal{C}$ and $\mathcal{S}$ are *internal* symmetries, hence $U_s$ cannot mix different sites of the photonic lattice.

Likewise, using the chiral symmetry relation (S40) (with $\tilde{U}_S$ in place of $U_S$), yields

$$\phi_S = \chi_S + \pi, \qquad \omega_0 = \bar{\omega}_c \tag{S45}$$

with $\langle v|U_S|v\rangle = e^{i\chi_S}$.

Thus, for either internal symmetry, the condition $\omega_0 = \bar{\omega}_c$ ensures that the total Hamiltonian $H$ enjoys the same internal symmetry as the lattice free Hamiltonian $H_B$.

The above argument makes clear that, if $H$ has one dressed BS occurring in the central bandgap, this is TR iff $\omega_0 = \bar{\omega}_c$. This condition yet is equivalent to stating that such a dressed BS is a VDS (the one associated with the photonic BS of lattice $B$ with energy $\omega_\psi = \bar{\omega}_c$).

## SM7. STABILITY OF THE VDS AGAINST DETUNING

Let $B$ be a translationally-invariant lattice and assume that BS admits an in-gap BS $|\psi\rangle$ of energy $\omega_\psi$. Then, if the atomic frequency is tuned such that $\omega_0 = \omega_\psi$, a VDS exists with energy $\omega_\Psi = \omega_0 = \omega_\psi$. Here, assuming weak coupling, we ask how sensitive is the VDS to the condition $\omega_0 = \omega_\psi$.

We thus introduce a small detuning between the atom and BS $|\psi\rangle$, $\Delta = \omega_0 - \omega_\psi$, which corresponds to a perturbation of the total Hamiltonian according to $H \to H + \Delta \sigma_+ \sigma_-$.

The unperturbed Hamiltonian can be written as

$$H = \omega_\psi |\Psi\rangle \langle \Psi| + \sum_m \omega_m |\Psi_m\rangle \langle \Psi_m|, \tag{S46}$$

where $|\Psi\rangle$ is the VDS [cf. Eq. (7)] and $|\Psi_m\rangle$ all the other single-photon dressed states.

Since $|\Psi\rangle$ is non-degenerate, we can apply standard second-order perturbation theory. Accordingly, the corrected dressed state $|\Psi_\Delta\rangle$ (such that $|\Psi\rangle_{\Delta=0} = |\Psi\rangle$ with $|\Psi\rangle$ the ideal VDS) is given by [cf. Eq. (7)]

$$|\Psi_\Delta\rangle = |\Psi\rangle + \Delta \cos\theta \sum_m \frac{\langle\Psi_m|e\rangle}{\omega_\psi - \omega_m} |\Psi_m\rangle + \Delta^2 \cos\theta \sum_m \frac{\langle\Psi_m|e\rangle}{\omega_\psi - \omega_m} \left( \sum_{m'} \frac{|\langle\Psi_{m'}|e\rangle|^2}{\omega_\psi - \omega_{m'}} - \frac{\cos^2\theta}{\omega_\psi - \omega_m} \right) |\Psi_m\rangle \,, \quad (S47)$$

and the corrected energy by

$$\omega_\Delta = \omega_\psi + \Delta |\cos\theta|^2 + \Delta^2 |\cos\theta|^2 \sum_m \frac{|\langle e|\Psi_m\rangle|^2}{\omega_\psi - \omega_m} \,. \quad (S48)$$

On the other hand, to first order in $g$ (we are assuming weak coupling), the unperturbed eigenstates $|\Psi_m\rangle$ can be expressed as

$$|\Psi_m\rangle = |\beta_m\rangle + g \frac{\langle v|\beta_m\rangle}{\omega_\psi - \omega_m} |e\rangle \,, \quad (S49)$$

with $|\beta_m\rangle$ single-photon eigenstates of $H_B$ such that $H_B |\beta_m\rangle = \omega_m |\beta_m\rangle$ and $V = g\,(b_v^\dagger \sigma_- + b_v \sigma_+)$ [cf. Eq. (1) in the main text].

Thus, so long as both $g$ and $\Delta$ are small compared to the distance in energy from the closer band, which is $\min_m |\omega_m - \omega_\psi|$, then up to first order the VDS wavefunction is insensitive to the detuning $\Delta$, only acquiring a small energy shift $\Delta |\cos\theta|^2$ [cf. Eq. (S48)].

### SM8. MANY-ATOM EFFECTIVE HAMILTONIANS MEDIATED BY IN-GAP VDS

#### SM8.1. Bound VDS in terms of the $H_B$'s normal modes

Let $\{|k\rangle\}$ be the single-photon eigenstates of $H_B$ such that $H_B |k\rangle = \omega_k |k\rangle$, where $k$ in the present subsection generically labels the $B$'s normal modes. These states can be used as a basis to expand $|\psi\rangle$ [cf. Eq. (3)] as $|\psi\rangle = \sum_k \psi_k |k\rangle$. Then Eq. (4) is equivalent to the coupled equations

$$\omega_0 \varepsilon + \sum_k g \langle v| k\rangle \psi_k = \omega_0 \varepsilon \,, \quad \omega_k \psi_k + g \langle k| v\rangle \varepsilon = \omega_0 \psi_k \,, \quad (S50)$$

Solving the latter equation for $\psi_k$ we get

$$\psi_k = \frac{g \langle k| v\rangle}{\omega_0 - \omega_k} \varepsilon \,. \quad (S51)$$

When the VDS (3) is bound, the normalization condition $|\varepsilon|^2 + \sum_k |\psi_k|^2 = 1$ must hold. Using (S51) and solving for $\varepsilon$ yields

$$\varepsilon = \frac{1}{\sqrt{1 + g^2 \sum_k \frac{|\langle k|v\rangle|^2}{(\omega_0 - \omega_k)^2}}} \,. \quad (S52)$$

Here, we assumed $\varepsilon \equiv |\varepsilon|$ (always possible by attaching to $|\Psi\rangle$ a suitable global phase factor). Replacing in (S51) we end up with

$$\tilde\psi_k = \frac{g \langle k| v\rangle}{\omega_0 - \omega_k} \frac{1}{\sqrt{1 + g^2 \sum_k \frac{|\langle k|v\rangle|^2}{(\omega_0 - \omega_k)^2}}} \,. \quad (S53)$$

Here, we added the tilde to recall that (S53) is unnormalized, thus different from that appearing in (7). The two are related as $\tilde\psi_k = e^{i\varphi} \sin\theta \psi_k$.

Eqs. (S52) and (S53) express a bound VDS as an explicit function of $g_k = g \langle k| v\rangle$, i.e., the coupling strength between the atom and mode $k$ of $B$ (note that it is $B$, not $B_v$). We point out that, for a general dressed BS which

is not necessarily also a VDS, this functional dependence is implicit in that in Eqs. (S52)-(S53) $\omega_0$ is replaced by the dressed-state energy which is a priori unknown when the dressed BS is not a VDS.

For $g = 0$ (zero coupling), Eqs. (S52) and (S53) yield $\varepsilon = 1$ and $\tilde\psi_k = 0$ as expected ($|\Psi\rangle = |e\rangle$). The next order of approximation is

$$\varepsilon \simeq 1, \ \tilde\psi_k \simeq \frac{g \langle k| v \rangle}{\omega_0 - \omega_k} \tag{S54}$$

(note that normalization is ensured to leading order). In terms of basis $\{|i\rangle\}$ of $B$ (real-space representation), using that $\tilde\psi_i = \sum_k \tilde\psi_k \langle i|k\rangle$, we get

$$\tilde\psi_i = g \sum_k \frac{\langle k|v\rangle \langle i|k\rangle}{\omega_0 - \omega_k}, \tag{S55}$$

which we recall that holds in the weak-coupling limit.

### SM8.2. Photonic lattice: many-atom effective Hamiltonian

Let now $B$ be a translationally-invariant photonic lattice, whose unit cell has $d$ cavities. Its free Hamiltonian is written in terms of normal modes as

$$H_B = \sum_{\mu,\mathbf{k}} \omega_{\mu,\mathbf{k}} \beta^\dagger_{\mu,\mathbf{k}} \beta_{\mu,\mathbf{k}} \tag{S56}$$

with $\mu$ a band index and $\mathbf{k}$ now standing for a (generally three-dimensional) wave vector. Denoting by $n$ the cell index, let $\mathbf{x}_{n,\alpha}$ with $\alpha = 1, ..., d$ be the (generally three-dimensional) position of the $\alpha$th cavity in cell $n$. When applied to the present lattice, Eq. (S55) thus specifically reads

$$\tilde\psi_{n,\alpha} = g \sum_{\mu,\mathbf{k}} \frac{\langle \mu, \mathbf{k}| n_v, \alpha_v\rangle \langle n, \alpha|\mu, \mathbf{k}\rangle}{\omega_0 - \omega_{\mu,\mathbf{k}}} \tag{S57}$$

(the atom is coupled to the $\alpha$th cavity of cell $n_v$). Here, $|\mu, \mathbf{k}\rangle = \beta^\dagger_{\mu,\mathbf{k}} |\text{vac}\rangle$, which due to translational invariance has the real-space Bloch-form form $|\mu, \mathbf{k}\rangle = \sum_{n,\alpha} c_{\mu,\mathbf{k},\alpha} e^{i\mathbf{k}\cdot\mathbf{x}_{n,\alpha}} |n, \alpha\rangle$. Therefore,

$$\tilde\psi^v_{n,\alpha} = g \sum_{\mu,\mathbf{k}} \frac{c^*_{\mu,\mathbf{k},\alpha_v} c_{\mu,\mathbf{k},\alpha} e^{i\mathbf{k}\cdot(\mathbf{x}_{n,\alpha}-\mathbf{x}_v)}}{\omega_0 - \omega_{\mu,\mathbf{k}}} \tag{S58}$$

where we set $\mathbf{x}_v = \mathbf{x}_{n_v,\alpha_v}$ and (in view of the many-atom generalization) added superscript $v$.

Consider next $N_a$ identical atoms indexed by $\nu = 1, ..., N_a$ with the $\nu$th atom coupled to cavity $(n_\nu, \alpha_\nu)$. The interaction Hamiltonian then reads

$$V = g \sum_\nu \left( b^\dagger_{n_\nu,\alpha_\nu} \sigma_{\nu-} + \text{H.c.} \right) = \sum_\nu \sum_{\mu,k} \left( g^\nu_{\mu,\mathbf{k}} \beta^\dagger_{\mu,\mathbf{k}} + \text{H.c.} \right) \tag{S59}$$

with

$$g^\nu_{\mu,\mathbf{k}} = g \, c^*_{\mu,\mathbf{k},\alpha_\nu} e^{-i\mathbf{k}\cdot\mathbf{x}_\nu}, \tag{S60}$$

where we set $\mathbf{x}_\nu = \mathbf{x}_{n_\nu,\alpha_\nu}$.

Let the atomic transition frequency $\omega_0$ lie well within a bangap of $H_B$. Thus, if $g$ is much smaller than the bandgap width $\Delta\omega_{\text{gap}}$, the atoms are far-detuned from all lattice modes $\beta_{\mu,\mathbf{k}}$. Then it can be shown in various ways [13, 14] that the photonic degrees of freedom can be adiabatically eliminated giving rise to the effective decoherence-free atom-atom interaction Hamiltonian

$$H_{\text{eff}} = \sum_{\nu\nu'} K_{\nu\nu'} \sigma_{\nu'+} \sigma_{\nu-} + \text{H.c.} \tag{S61}$$

with the second-order coupling strengths given by

$$K_{\nu\nu'} = \tfrac{1}{2} \sum_{\mu,\mathbf{k}} \frac{g_{\mu,\mathbf{k}}^{\nu*} g_{\mu,\mathbf{k}}^{(\nu')}}{\omega_0 - \omega_{\mu,\mathbf{k}}} = \frac{g^2}{2} \sum_{\mu,\mathbf{k}} \frac{c_{\mu,\mathbf{k},\alpha_{\nu'}}^* c_{\mu,\mathbf{k},\alpha_v} e^{i\mathbf{k}\cdot(\mathbf{x}_\nu - \mathbf{x}_{\nu'})}}{\omega_0 - \omega_{\mu,\mathbf{k}}} \,, \tag{S62}$$

where in the last identity we used (S60). Upon comparison with (S58), we thus end up with

$$K_{\nu\nu'} = \tfrac{1}{2} g \tilde{\psi}_{n_\nu,\alpha_\nu}^{\nu'} = \tfrac{1}{2} g \sin\theta_{\nu'} e^{i\varphi_{\nu'}} \, \psi_{n_\nu,\alpha_\nu}^{(\nu')} \tag{S63}$$

where in the last identities we introduced $|\psi\rangle$ (normalized) using (7) and added a subscript $\nu'$ to the angles (8) since these depend on the position of the $\nu'$th atom. Finally, in order to ensure consistency with Eqs. (7)-(8) (where $\theta$ and $\varphi$ are at all orders in $g$), we must approximate $\theta = \arctan|\eta| \simeq |\theta|$. Thus, recalling that $\varphi = \arg \eta$, we get $\sin\theta e^{i\varphi} \simeq \theta e^{i\varphi} = \eta$ so as to end up with

$$K_{\nu\nu'} = \tfrac{1}{2} g \eta_{\nu'} \, \psi_{n_\nu,\alpha_\nu}^{(\nu')} = -\frac{g^2}{2\langle\nu|H_B|\psi^{(\nu')}\rangle} \, \psi_\nu^{(\nu')} \,, \tag{S64}$$

where for brevity we set $\nu \equiv (n_\nu, \alpha_\nu)$.

## SM9. VDS IN THE PHOTONIC SSH MODEL

When $v = 1$ (atom coupled to cavity $a$ in cell $n=1$) and for $\delta > 0$, the photonic wavefunction is non-zero only on even sites (i.e., cavities $b$) and reads (see, e.g., Ref. [15])

$$\psi_{2n} = \frac{2\sqrt{\delta}}{1+\delta} \left(\frac{\delta-1}{\delta+1}\right)^{N-n} \tag{S65}$$

with $n = 2, ..., N$. For $\delta < 0$, this must be mirror-reflected around $v = 1$ making the simultaneous replacement $\delta \to -\delta$. Plugging this into (8) directly yields the dressed BS (7) with $\theta = \arctan[g/(2J\sqrt{\delta}\,)]$ and $\varphi = 0$.

## SM10. THEOREM FOR 1D LATTICES: BS OF $B_v$ FROM EDGE STATES UNDER OPEN BCS

Consider a 1D photonic lattice $B$. Note that the lattice under *open* BCs is obtained from the translationally-invariant lattice $B$ by removing an entire cell (instead of a single cavity as in the definition of $B_v$). Here, we derive a condition allowing to deduce both the existence and wavefunction of the photonic BS of $B_v$ from in-gap edge states under open BCs (if any).

Let $B$ have $N$ unit cells, each with $d$ cavities. Then the most general free Hamiltonian of the lattice can be written as

$$\mathcal{H}_\lambda^N = \sum_{n=1}^{N} \sum_{\alpha,\alpha'=1}^{d} b_{n,\alpha}^\dagger h_{\alpha\alpha'} b_{n,\alpha'} + \sum_{r=1}^{R} \sum_{n=1}^{N} \sum_{\alpha,\alpha'=1}^{d} \left( b_{n,\alpha}^\dagger J_{\alpha\alpha'}^r b_{n+r,\alpha'} + \text{H.c.} \right) \tag{S66}$$

Here, the $d \times d$ Hermitian matrix $h_{\alpha\alpha'}$ specifies the intra-cell cavity couplings (off-diagonal entries) and on-site cavity frequencies (diagonal entries), while the (generally non-symmetric) $d \times d$ matrix $J_{\alpha\alpha'}^r$ contains all the inter-cell cavity-cavity couplings with range $r = 1, 2, ..., R$ (for nearest-neighbor cells, $r = 1$; $R$ is the maximum range). Here, we conveniently introduced the notation $\mathcal{H}_\lambda^N$, where the superscript is the number of cells while the subscript $\lambda = P, O, v$ specifies the BCs with $P$ standing for periodic BCs (translationally-invariant lattice $B$), $O$ for the lattice subject to open BCs ($B$ without an entire cell) and $v$ for lattice $B_v$. Thus, to connect with the main-text notation, $H_B = \mathcal{H}_P^N$, $H_{B_v} = \mathcal{H}_v^N$.

We note that, if $R > 1$ one can always redefine the lattice unit cell such such that it contains $Rd$ cavities. Accordingly, without loss of generality, henceforth we focus on nearest-neighbor inter-cell couplings and thus set $R = 1$ (it is understood that, if $R > 1$, $d$ must be intended as $Rd$). It is convenient to introduce a vector-matrix formalism allowing to rewrite (S66) as

$$H_\lambda^N = \sum_{n=1}^{N} \boldsymbol{\varphi}_n^\dagger \cdot \mathbf{h} \cdot \boldsymbol{\varphi}_n + \sum_{n=1}^{N} \left( \boldsymbol{\varphi}_n^\dagger \cdot \mathbf{J} \cdot \boldsymbol{\varphi}_{n+1} + \text{H.c.} \right) \tag{S67}$$

with $\mathbf{h}_0$ and $\mathbf{h}_1$ the matrices corresponding to rates $h_{ij}$ and $J_{ij}^{r=1}$, respectively, and where $\boldsymbol{\varphi}_n = \begin{pmatrix} b_{n,1} & \ldots & b_{n,d} \end{pmatrix}^T$.

Consider now a vacancy on site $\alpha_v$ of cell $n = 0$. It is convenient to define $\tilde{\boldsymbol{\varphi}}_0 = (\mathbb{1}_d - \mathbf{P}) \cdot \boldsymbol{\varphi}_0$, where $\mathbf{P}$ is the $d$-dimensional projector on the vacancy site (inside the unit cell); thus $\tilde{\boldsymbol{\varphi}}_0$ is simply $\boldsymbol{\varphi}_0$ with the $\alpha_v$th component set to zero. The free Hamiltonian of $B_v$, with $B$ having $N+1$ cells, can thus be written as

$$H_v^{(N+1)} = H_O^N + \tilde{\boldsymbol{\varphi}}_{N+1}^\dagger \cdot \mathbf{h} \cdot \tilde{\boldsymbol{\varphi}}_{N+1} + \tilde{V}, \tag{S68}$$

where

$$\tilde{V} = \tilde{\boldsymbol{\varphi}}_0^\dagger \cdot \left( \mathbf{J} \cdot \boldsymbol{\varphi}_1 + \mathbf{J}^\dagger \cdot \boldsymbol{\varphi}_N + \text{H.c.} \right) \tag{S69}$$

is the coupling Hamiltonian between all cavities of cell $n = 0$ but $\alpha_v$ (vacancy site) and lattice cells $n = 1, ..., N$ (i.e, $B$ under open BCs).

Assume now that $B$ under open BCs (cells $n = 1, ..., N$) admits $\mathcal{N}_e$ degenerate edge states of energy $\omega_e$, which we call $|\mathcal{E}^s\rangle = \sum_{n=1}^{N} \boldsymbol{\varphi}_n^\dagger \cdot \boldsymbol{\mathcal{E}}_n^s |\text{vac}\rangle$ with $s = 1, ..., \mathcal{N}_e$ (here $\boldsymbol{\mathcal{E}}_n^s$ is a $d$-dimensional row vector). Being these localized, $\omega_e$ lies within a bandgap (note that $B$, $B$ under open BCs and $B_v$ share the same bands and bandgaps). Consider now a linear combination of these edge states, $|\psi\rangle = \sum_{s=1}^{\mathcal{N}_e} \gamma_s |\mathcal{E}^s\rangle$. The condition in order for $|\psi\rangle$ to be an eigenstate of $H_v^{(N+1)}$ with eigenvalue $\omega_e$ is

$$\tilde{V} |\psi\rangle = 0, \tag{S70}$$

where we used that $H_O^N |\mathcal{E}^s\rangle = \omega_e |\mathcal{E}^s\rangle$ and $\langle 0|b_{0,\alpha}|\mathcal{E}^s\rangle = 0$ for $\alpha = 1, ..., d$.

Eq. (S70) is a linear system of $d - 1$ equations in the $\mathcal{N}_e$ unknowns $\{\gamma_s\}$. This has $\mathcal{N}_e - d + 1$ non-trivial solutions. Hence, if $\mathcal{N}_e = d$ there is only one non-trivial solution, which is a bound state of $B_v$. This completes the proof.

## SM11. VDS IN THE PHOTONIC CREUTZ-LADDER MODEL

In this section, we consider $B$ to be a photonic Creutz-ladder model and show that, when the bandgap is open, $B_v$ admits a photonic BS (hence a corresponding VDS occurs). This task is carried out by applying the theorem in the last section by first deriving the edge states of $B$ under open BCs through the methods introduced in Refs. [16, 17].

### SM11.1. Edge States of $B$ under open BCs

The free Hamiltonian of the Creutz model with open BCs is

$$\mathcal{H}_O^{(N)} = -2mJ \sum_{n=1}^{N} \left( a_n^\dagger b_n + \text{H.c.} \right) + J \sum_{n=1}^{N-1} \left[ e^{i\alpha} a_n^\dagger a_{n+1} + e^{-i\alpha} b_n^\dagger b_{n+1} + a_n^\dagger b_{n+1} + b_n^\dagger a_{n+1} + \text{H.c.} \right], \tag{S71}$$

where $a_n$ and $b_n$ are ladder operators corresponding to cavities $a$ and $b$ of cell $n$ [see Fig. 2(e)] and where we set $\omega_c = 0$ (which does not affect the calculation). Column vector $\boldsymbol{\varphi}_n$ and matrices $\mathbf{h}_{0,1}$ in Eq. (S67) in this case thus read

$$\boldsymbol{\varphi}_n = \begin{pmatrix} a_n \\ b_n \end{pmatrix}, \qquad \mathbf{h}_0 = -2Jm\, \sigma_x, \qquad \mathbf{h}_1 = J \begin{pmatrix} e^{i\alpha} & 1 \\ 1 & e^{-i\alpha} \end{pmatrix} \tag{S72}$$

($\sigma_{x,y,z}$ are the usual Pauli matrices).

Consider first the generalized Bloch Hamiltonian [16, 17]

$$\mathcal{H}(z) = \mathbf{h}_0 + z\mathbf{h}_1 + z^{-1}\mathbf{h}_1^\dagger = -J \begin{pmatrix} -e^{i\alpha}z - e^{-i\alpha}z^{-1} & 2m - (z + z^{-1}) \\ 2m - (z + z^{-1}) & -e^{-i\alpha}z - e^{i\alpha}z^{-1} \end{pmatrix}, \tag{S73}$$

where we set $z = e^{ik}$ with $k$ complex. To find the localized states of $\mathcal{H}_O^{(N)}$, it is enough to find the roots of the characteristic polynomial

$$P(E, z) = \det \left[ \mathcal{H}(z) - E\mathbb{1}_2 \right] = 2(E \cos \alpha - 2Jm)(z + z^{-1}) + E^2 - 4J^2m^2 - 4J^2 \sin^2 \alpha, \tag{S74}$$

which are

$$z_\pm = z_\pm(E) = \omega_E \pm \sqrt{\omega_E^2 - 1} \quad \text{with} \quad \omega_E = \frac{4J^2 m^2 + 4J^2 \sin^2\alpha - E^2}{4(E\cos\alpha - 2Jm)} \tag{S75}$$

(note that $z_- = 1/z_+$; the notation stresses that $z_\pm$ are functions of $E$).

Let $\mathcal{H}_\infty$ be the Hamiltonian obtained from (S71) by replacing each sum with $\sum_{n=-\infty}^{\infty}$. Then a generic eigenstate of $\mathcal{H}_\infty$ can be written as

$$|\chi_\pm\rangle = \sum_{n=-\infty}^{+\infty} z_\pm^n \, \boldsymbol{\varphi}_n^\dagger \cdot \mathbf{u}(z_\pm) \, |\text{vac}\rangle \,, \tag{S76}$$

where $\mathbf{u}(z_\pm)$ is the (two-dimensional) eigenvector of $\mathcal{H}(z)$ with eigenvalue $E$.

Next, to work out the eigenstates of $\mathcal{H}_O^{(N)}$, we make the ansatz

$$|\mathcal{E}\rangle = \mathbb{P}^N \left( c_+ |\chi_+\rangle + c_- |\chi_-\rangle \right) \,, \tag{S77}$$

where $\mathbb{P}^N = \sum_{n=1}^{N} a_n^\dagger |\text{vac}\rangle\langle\text{vac}| a_n + b_n^\dagger |\text{vac}\rangle\langle\text{vac}| b_n$ is the projector on all the sites that belong to the lattice. In order to be an eigenstate of $\mathcal{H}_O^{(N)}$, $|\mathcal{E}\rangle$ must satisfy the eigenvalue equation in particular on the lattice boundary. This corresponds to enforcing the conditions

$$\sum_{i=\pm} c_i \, \langle\text{vac}| \, a_n (\mathcal{H}_O^{(N)} - E\mathbb{1}) |\chi_i\rangle = 0 \,, \quad \sum_{i=\pm} c_i \, \langle\text{vac}| \, b_n (\mathcal{H}_O^{(N)} - E\mathbb{1}) |\chi_i\rangle = 0 \quad \text{with} \ n = 1, N \,, \tag{S78}$$

which is a linear system of four equations in the unknowns $c_\pm$ (each equation corresponds to site $a$ or $b$ of cell $n=1$ or $n=N$). The solutions for $z_\pm$ are those values of $z_\pm$ such that system (S78) admits non-trivial solutions. Based on Eq. (S76), note that if $|z_\pm| \neq 1$ then the corresponding eigenstate is localized close to one of the two lattice edges, the left edge if $|z_\pm| < 1$, the right one if $|z_\pm| > 1$ (left and right edged correspond to cells $n = 1/n = N$, respectively).

After some manipulations, in the thermodynamical limit $N \to \infty$, we end up with [recall Eq. (S75)]

$$E = 2m \cos\alpha \, J \,, \quad z_\pm = m^{\pm 1} \,. \tag{S79}$$

Plugging these into (S77) [recall Eq. (S76)], we get the pair of edge states

$$|\mathcal{E}^L\rangle = \sqrt{\tfrac{1-m^2}{2}} \sum_{n=1}^{N} m^{n-1} \left( e^{i\frac{\alpha}{2}} a_n^\dagger - e^{-i\frac{\alpha}{2}} b_n^\dagger \right) |\text{vac}\rangle \,, \quad |\mathcal{E}^R\rangle = \sqrt{\tfrac{1-m^2}{2}} \sum_{n=1}^{N} m^{N-n} \left( e^{-i\frac{\alpha}{2}} a_n^\dagger - e^{i\frac{\alpha}{2}} b_n^\dagger \right) |\text{vac}\rangle \,, \tag{S80}$$

with subscript $L$ ($R$) indicating whether the state is localized close to the left (right) edge (the wavefunction modulus decays from right to left in the case of $\mathcal{E}^R$ and from left to right for $\mathcal{E}^L$).

### SM11.2. Bound state of $B_v$

In light of the theorem in Section SM10, the knowledge of edge states (S80) ensure the existence of a BS of $B_v$. To work out the corresponding wavefunction, we just impose (S70) using (S80), obtaining

$$|\psi_{v=a}\rangle = \tfrac{1}{\sqrt{2}} \left( |\mathcal{E}^R\rangle + e^{-i\alpha} |\mathcal{E}^L\rangle \right) = \frac{\sqrt{1-m^2}}{2} e^{-i\frac{\alpha}{2}} \sum_{n=2}^{N} \left[ \left( e^{i\alpha} m^{n-2} + e^{-i\alpha} m^{N-n} \right) a_n^\dagger - \left( m^{n-2} + m^{N-n} \right) b_n^\dagger \right] \,, \tag{S81}$$

$$|\psi_{v=b}\rangle = \tfrac{1}{\sqrt{2}} \left( |\mathcal{E}^R\rangle + e^{i\alpha} |\mathcal{E}^L\rangle \right) = \frac{\sqrt{1-m^2}}{2} e^{i\frac{\alpha}{2}} \sum_{n=2}^{N} \left[ \left( m^{n-2} + m^{N-n} \right) a_n^\dagger - \left( e^{-i\alpha} m^{n-2} + e^{i\alpha} m^{N-n} \right) b_n^\dagger \right] \tag{S82}$$

with subscript $v = a, b$ indicating whether the atom is coupled to cavity $a$ or $b$.

## SM12. HALDANE MODEL

### SM12.1. Existence of a VDS

Here we will prove that a *BS of $B_v$* in the Haldane model occurs if and only if the model lies within regions I and II of the parameters space [see ]Fig. 3(a) in the main text]. Thus, a VDS can occur only within the same regions.

Recall from Section SM5 that a BS of $B_v$ exists iff $\langle G_B(z)\rangle_v$ admits a root $z = \omega_p$ within the bandgap. Since $\langle G_B(z)\rangle_v$ is analytic and monotonic, this occurs iff $\langle G_B(z)\rangle_v$ takes opposite signs at the band-gap edges.

In the reciprocal space, the Haldane model can be expressed as

$$H_B = \iint_{\mathbb{T}^2} d^2\boldsymbol{k}\, \boldsymbol{\varphi}_{\boldsymbol{k}}^\dagger \cdot \mathcal{H}_{\boldsymbol{k}} \cdot \boldsymbol{\varphi}_{\boldsymbol{k}} \qquad \boldsymbol{\varphi}_{\boldsymbol{k}} = \frac{1}{2\pi}\iint_{\mathbb{T}^2} d^2\boldsymbol{k}\, e^{-i\pi\boldsymbol{k}\cdot\boldsymbol{x}}\boldsymbol{\varphi}_{\boldsymbol{x}}, \qquad \boldsymbol{\varphi}_{\boldsymbol{x}} = \begin{pmatrix} a_{\boldsymbol{x}} \\ b_{\boldsymbol{x}} \end{pmatrix}, \tag{S83}$$

$$\mathcal{H}_{\boldsymbol{k}} = J\sum_{i=1}^{3}\left\{ 2t\cos\phi\cos(\boldsymbol{k}\cdot\boldsymbol{d}_i)\mathbb{1} + \cos(\boldsymbol{k}\cdot\boldsymbol{u}_i)\sigma_x + \sin(\boldsymbol{k}\cdot\boldsymbol{u}_i)\sigma_y + [m - 2t\sin\phi\sin(\boldsymbol{k}\cdot\boldsymbol{d}_i)]\sigma_z\right\}, \tag{S84}$$

where $a_{\boldsymbol{x}}$ ($b_{\boldsymbol{x}}$) is the annihilation operator of cavity $a$ ($b$) on the unit cell specified by vector $\boldsymbol{x}$, $\boldsymbol{u}_1 = (0,1)^T$, $\boldsymbol{u}_2 = (-\sqrt{3}/2, -1/2)^T$, $\boldsymbol{u}_3 = (\sqrt{3}/2, -1/2)^T$ and $\boldsymbol{d}_i = \boldsymbol{u}_i - \boldsymbol{u}_{i-1}$ ($i = 1\ldots 3 \mod 3$).

For $\Delta\omega_{\text{gap}} \ll J$, the low-energy physics is well described by the 2D-Dirac Hamiltonian

$$h_{\pm}(\boldsymbol{\kappa}) \simeq \omega_{\text{mid}}\mathbb{1} \pm \left[ M_{\pm}c^2\sigma_z + c(-\kappa_x\sigma_x \pm \kappa_y\sigma_y)\right] \quad \text{with } M_{\pm} = J\frac{3\sqrt{3}t\sin\phi\pm m}{c^2}, \quad c = \frac{3J}{2}, \quad \omega_{\text{mid}} = -3Jt\cos\phi$$

obtained by expanding (S84) in momenta $\boldsymbol{\kappa} = \boldsymbol{k} \pm \boldsymbol{K}$ around the two Dirac points $\pm\boldsymbol{K}$, where $\boldsymbol{K} = \left(\frac{4\pi}{3\sqrt{3}1},0\right)$.

Accordingly, the lattice Green function is given by $G_B(z) = (z - H_B)^{-1} \simeq \sum_{\pm}\iint d^2\boldsymbol{\kappa}(\omega - h_{\pm}(\boldsymbol{\kappa}))^{-1}\boldsymbol{\varphi}_{\boldsymbol{\kappa}\pm\boldsymbol{K}}^\dagger\boldsymbol{\varphi}_{\boldsymbol{\kappa}\pm\boldsymbol{K}}$, which yields

$$\langle G_B(z)\rangle_v \simeq -\frac{\pi}{c^2}\sum_{\pm}(z' \pm M_{\pm}c^2\langle\sigma_z\rangle_v)\ln\left|1 + \frac{c^2\pi^2}{M_{\pm}^2c^4 - z'^2}\right| \qquad \text{with } z' = z - \omega_{\text{mid}} \in \mathbb{R}, \tag{S85}$$

with $v = a$ or $v = b$, in which cases $\langle\sigma_z\rangle_v = 1$ and $\langle\sigma_z\rangle_v = -1$, respectively.

Without loss of generality, we assume $m > 0$ and $0 < \phi < \pi$ such that $|M_+| > |M_-|$ and the bandgap edges are at $z' = \pm M_-c^2$. In regions I and II [see Fig. 3(a) in the main text], $|m| < 3\sqrt{3}t|\sin\phi|$, hence $M_- > 0$ so that the upper and lower bandgap edges lie at $M_-c^2$ and $-M_-c^2$, respectively. Thereby, the Green function takes the following values on the bandgap edges

$$\text{lower edge:} \qquad \lim_{z'\to(-M_-c^2)^+}\langle G_B(z')\rangle_v \simeq \begin{cases} +\infty & \text{for } v = a \\ +\pi\ln\left|1 + \frac{(\pi/a)^2}{(M_+^2 - M_-^2)c^2}\right|(M_- + M_+) > 0 & \text{for } v = b \end{cases} \tag{S86}$$

$$\text{upper edge:} \qquad \lim_{z'\to(+M_-c^2)^-}\langle G_B(z')\rangle_v \simeq \begin{cases} -\pi\ln\left|1 + \frac{(\pi/a)^2}{(M_+^2 - M_-^2)c^2}\right|(M_- + M_+) < 0 & \text{for } v = a \\ -\infty & \text{for } v = b \end{cases} \tag{S87}$$

Therefore, no matter whether the atom sits on $a$ or $b$, $\langle G_B(z')\rangle_v$ changes sign across the bandgap, entailing $\langle G_B(\omega_p)\rangle_v = 0$ for some $\omega_p$ within the gap.

On the other hand, for $|m| > 3\sqrt{3}t|\sin\phi|$ (i.e., outside regions I-II), $M_- < 0$ and the upper (lower) band-gap edge occurs at $z' = -M_-c^2$ ($z' = M_-c^2$). Then

$$\text{lower edge:} \qquad \lim_{z'\to(+M_-c^2)^+}\langle G_0(z')\rangle_v \simeq \begin{cases} -\pi\ln\left|1 + \frac{(\pi/a)^2}{(M_+^2 - M_-^2)c^2}\right|(M_- + M_+) < 0 & \text{for } v = a \\ +\infty & \text{for } v = b \end{cases} \tag{S88}$$

$$\text{upper edge:} \qquad \lim_{z'\to(-M_-c^2)^-}\langle G_0(z')\rangle_v \simeq \begin{cases} -\infty & \text{for } v = a \\ +\pi\ln\left|1 + \frac{(\pi/a)^2}{(M_+^2 - M_-^2)c^2}\right|(M_- + M_+) > 0 & \text{for } v = b \end{cases} \tag{S89}$$

Thus, contrary to the previous case, $\langle G_0(z')\rangle_v$ does *not* change sign accross the bandgap wherein it cannot vanish.

### SM12.2. Additional remarks on topological protection of the VDS for $\phi = \pm\pi/2$ and $m = 0$.

As mentioned in the main text, a VDS inherits its topological features (if any) from the BS of $B_v$, i.e., the BS induced by a (zero-dimensional) vacancy.

In the Haldane model, a topologically robust BS around a vacancy is guaranteed only for $\phi = \pm\pi/2$ and $m = 0$. Indeed, according to the topological classification in Ref. [8], a zero-dimensional defect in a 2D model may seed around it a TR BS only when the model lies in a suitable Atland-Zirnbauer class [9]. Classes are identified by the occurrence of time-reversal, particle-hole and chiral symmetry, or the absence thereof. The Haldane model *generally* belongs to class A, the class of models lacking any of the above symmetries (although some symmetry can occur on special points of the parameter space as discussed shortly). According to Ref. [8], class-A models may feature one-dimensional topologically-protected states. In the Haldane model, these states are the well known chiral edge states which appear in the topological phases I and II of Fig. 3(a) in the main text. Note that, according to the classification of Ref. [8], models within class A do not admit zero-dimensional TR BS. Yet, it turns out that, for the special values $\phi = \pm\pi/2$ and $m = 0$, the Haldane model does possess particle-hole symmetry, so that the model falls within a different class, namely class D. Two-dimensional class-D models may have zero-dimensional TR BS, whenever a suitable $Z_2$ topological invariant [7] acquires a non-vanishing value. This is indeed the case in the instance of Fig. 3(b) [white dot of Fig. 3(a)]. The existence of these TR BS in the Haldane model is indeed analytically proven in Ref. [18], where moreover their topological protection is numerically confirmed. Note that a BS may still exist within the whole phases I and II [see Fig. 3(a) in the main text]: however, due to lack of particle-hole symmetry their topological protection is not guaranteed except on the special point $\phi = \pm\pi/2$ and $m = 0$ [18].

### SM12.3. Dressed bound states that are not VDS

Fig. 3(b) shows the photon current density (CD) of the topologically robust VDS for $m = 0$, $\pi/2$ and $t = 0.1$. In this case, the CD is highly picked, reaching a maximum value of 0.04.

There are points of the phase diagram of Fig. 3(a) outside the regions I and II, where the gap $\Delta\omega_{\text{gap}}$ coincides with the value $\Delta_0$ assumed at $\phi = \pm\pi/2$ and $m/t = 0$, and all other parameters unchanged. If one takes $\omega = \omega_{\text{mid}}$ and $g$ small enough, an in-gap dressed BS arises (which is not a VDS) in which the localisation of the photon CD is quantitatively similar to the one in the TR VDS of Fig. 3(b).

This dressed BS may still display a CD pinned around $v$ whose magnitude, however, is several orders of magnitude smaller than in Fig. 3(b). For example, for the set of dressed BS occurring for $m = \Delta_0 - 3\sqrt{3}|\sin\phi|$, $\phi \in [-\pi, \pi]$, $t = 0.1$ and $g = 0.01$, for which $\Delta\omega_{\text{gap}} = \Delta_0$, we numerically observe a maximum CD that is at least six order of magnitudes smaller than the VDS in Fig. 3(b). This is based on exact numerical diagonalization of the Hamiltonian using a lattice of $30 \times 30$ unit cells and the CD formula [19]

$$\boldsymbol{j}(\boldsymbol{x}_j) = \sum_k (\boldsymbol{x}_j - \boldsymbol{x}_k)\Im\Big(\langle\Psi|\boldsymbol{x}_j\rangle\langle\boldsymbol{x}_j|H|\boldsymbol{x}_k\rangle\langle\boldsymbol{x}_k|\Psi\rangle\Big) \tag{S90}$$

where $\boldsymbol{x}_j$ is the position of the $j$-th site on the lattice.



\end{document}